\documentclass[%
superscriptaddress,
preprint,
showpacs,preprintnumbers,
 showkeys,
 amsmath,amssymb,
 aps,
 prc,
]{revtex4-1}

\usepackage{latexsym}
\usepackage{psfrag}
\usepackage{graphics}
\usepackage{graphicx}
\usepackage{makeidx}
\usepackage{color}
\usepackage{amsfonts}
\usepackage{dcolumn} % Align table columns on decimal point
\usepackage{bm} % bold math
\usepackage{setspace,supertabular}
\usepackage{lineno}
 \usepackage{subfigure}
\usepackage{appendix}
\usepackage{graphicx}% Include figure files
\usepackage{dcolumn}% Align table columns on decimal point
\usepackage{bm}% bold math
\begin{document}

\newcommand{\ket}[1]{{\vert #1 \rangle}}
\newcommand{\bra}[1]{{\langle #1 \vert}}
\newcommand{\brak}[2]{{\langle #1\vert #2 \rangle}}
\def\sls #1{\rlap/\kern  -  .1em #1}
\newcommand{\di}{\displaystyle}
\renewcommand{\vec}[1]{\mathbf{#1}}

\title{Elastic Differential Cross Sections for Space Radiation Applications}% Force line breaks with \\

\author{Charles M. Werneth}
\affiliation{%
 NASA Langley Research Center, 2 West Reid Street, Hampton, VA 23681
}
\author{Khin M. Maung}%
\affiliation{%
The University of Southern Mississippi, 118 College Drive, Box 5046, Hattiesburg, MS 39406 
}
\author{William~P.~Ford}%
\affiliation{%
The University of Southern Mississippi, 118 College Drive, Box 5046, Hattiesburg, MS 39406 
}
\author{John W. Norbury}
\affiliation{%
 NASA Langley Research Center, 2 West Reid Street, Hampton, VA 23681
}
\author{Michael D. Vera}%
\affiliation{%
The University of Southern Mississippi, 118 College Drive, Box 5046, Hattiesburg, MS 39406 
}

\date{\today}

\begin{abstract}
The eikonal, partial wave (PW) Lippmann-Schwinger, 
and three-dimensional Lippmann-Schwinger (LS3D) methods are compared for nuclear reactions that are relevant for space radiation applications.  
Numerical convergence of the eikonal method is readily achieved when exact formulas of the optical potential are used for light nuclei ($A \le 16$), 
and the momentum-space representation of the optical potential is used for heavier nuclei. 
The PW solution method is known to be numerically unstable for systems that require a large number of partial waves, and, as a result, the LS3D method is employed.  The effect of relativistic kinematics is studied with the PW and LS3D methods and is compared to eikonal results. 
It is recommended that the LS3D method be used for high energy nucleon-nucleus reactions and nucleus-nucleus reactions at all energies because of its rapid numerical convergence and stability.
%\begin{description}
%\end{description}
\end{abstract}

%Use showkeys class option if keyword
\maketitle

\section{Introduction}

The space radiation environment is composed of solar particle emissions and ions produced from supernovae distributed throughout the galaxy \cite{Ackermann,Benton}. Solar particle events, including both coronal mass ejections and solar flares, are composed of mostly protons with energies that can exceed several hundred MeV. Galactic cosmic rays originate from the shock waves of supernovae and consist of protons and heavier ions with energies that reach hundreds of GeV per nucleon. Radiation transport codes are used to describe the transport of ions, and secondary particles produced from nuclear collisions, from the space radiation environment through shielding materials. Space radiation transport codes require cross sections for the numerous nuclear reactions that occur as a result of collisions of nuclei in the space radiation environment with nuclei in the shield.  NASA's deterministic transport code, HZETRN \cite{Slaba1,Slaba2,NASARP1257}, currently transports all ions up to nickel---where, thereafter, incident particle fluxes are negligible \cite{Simpson}---with energies that extend from MeV to hundreds of GeV per nucleon through shielding materials. Efficient, accurate codes are needed for the computation of nuclear cross sections due to the large number of nuclear reactions that occur at these energies.

The Lippmann-Schwinger (LS) equation is an expression for the scattering transition amplitude \cite{Joachain}.  Scattering amplitudes can be obtained by either solving the LS equation or by employing some approximation, such as the eikonal method. The elastic differential cross section is computed from the absolute square of the scattering amplitude, and the total cross section is related to the imaginary part of the forward scattering amplitude. The elastic cross section is obtained by performing the angular integration of the elastic differential cross section, and the reaction cross section is found from the difference between the total and elastic cross sections.

The input into the elastic scattering equation is the optical potential, which can be expressed in an infinite series of nucleon-nucleon (NN) transition amplitudes, $t_{\rm NN}$. If the transition matrix is written for ground states of the projectile and target, then, in the factorization approximation, the optical potential is proportional to $t_{\rm NN}$ and the nuclear densities of the projectile and target \cite{Pickle,Pickle2,Wolfe}. The model of $t_{\rm NN}$ used in the present work is parameterized to NN total cross sections, slope parameters, and the real to imaginary ratios of the transition amplitude. Nuclear charge density distributions are obtained from electron scattering experiments \cite{DeVries1,DeVries2}. Matter densities of nuclei are found from nuclear charge densities by factoring out the charge distribution of the proton. The internal charge structure of the proton is not taken into account in this analysis;  instead, nucleons are treated as point particles. Harmonic well densities are typically used for lighter nuclei because of the Gaussian-like decay of the nuclear charge density as a function of radial distance. Wood-Saxon densities, also known as two-parameter and three-parameter Fermi densities, are better suited for heavier nuclei, where the nuclear charge density is relatively constant before decreasing to zero at larger radial distances.

The two most common ways of solving the LS equation are to use the eikonal approximation or the method of partial wave (PW) decomposition \cite{Joachain}. The eikonal approximation was first introduced by Moliere and systematically developed by Glauber in the treatment of many-body nuclear reactions with a quantum collision theory of composite objects \cite{Joachain, Miller}.  The eikonal approximation can be derived by assuming high energy and small angle scattering, which leads to a linearized propagator in the LS equation from which the eikonal scattering wave function may be obtained \cite{Joachain}. The scattering amplitude is determined from eikonal phase factor, which is a function of the the optical potential \cite{NASARP1257,Townsend, Wilson1974,CJP1981,CJP1982, CJP1983}. 

Besides being an approximation, a drawback of the eikonal approximation is that it may be numerically inefficient for the evaluation of the cross sections for a given optical potential. In the position-space representation, the optical potential, $U(\vec r)$, is given by a 6-dimensional integration for heavy ion collisions \cite{NASARP1257}. Therefore,
the eikonal phase factor depends on a 6-dimensional integral in the position representation of the optical potential and an additional integration variable over a coordinate in the scattering plane. The numerical integration over 7-dimensions in the position space representation is inefficient when an analytic expression of the optical potential is not known. It is desirable to use exact formulas for the optical potential when analytic expressions of the optical potential can be found. The current work implements expressions of the optical potential for nucleon-nucleus (NA) and nucleus-nucleus (AA) scattering utilizing harmonic well nuclear matter densities for light nuclei ($A \le 16$), and the optical potential is expressed in momentum space for cases where no analytic expression can be found ($A>16$) \cite{NASATP2014}.

The LS equation may also be solved via the method of partial wave decomposition \cite{Joachain,Werneth_Finite}, where the transition amplitude is expanded in an infinite series of functions of relative momenta and angular dependent spherical harmonics or Legendre polynomials. 
After integrating over the angular dependence, the transition amplitude is solved for a given partial wave.
Once the partial wave solutions are found, the full solution for the transition amplitude is found by re-summing the series, which is terminated when some pre-defined tolerance of precision is reached.

The PW method is known to become numerically unstable for reactions that require many partial waves \cite{Joachain}, which is not only limited to high energy NA reactions (GeV/n) but also includes AA reactions at relatively low energy per nucleon (hundreds of MeV/n). The numerical instability can be traced back to highly oscillating Legendre polynomials in the PW expansion and large on-shell momenta for elastic reactions, where contributions to the transition amplitude tend to be localized.  

Although there are numerical limitations associated with the PW method, the full three-dimensional Lippmann-Schwinger (LS3D) solution method circumvents the necessity of using highly oscillating Legendre polynomials \cite{LS3D_1,LS3D_2,LS3D_3,LS3D_4,LS3D_3body}.  Most of the LS3D studies have consisted of NN interactions \cite{LS3D_1,LS3D_2,LS3D_4} with the exception of Rodriguez--Gallardo et al. \cite{LS3D_3} who studied NA and AA reactions at relatively low energies and Liu et al. \cite{LS3D_3body} who studied three-body reactions. This demonstrates the validity of the method and can be compared to results generated with the PW method since few partial waves are needed for such reactions. In the present work, the LS3D method is compared to the PW and eikonal methods for NA and AA reactions with energies extending from 150 MeV/n to 20 GeV/n. 

The eikonal method is a non-relativistic approximation; however, when energies become sufficiently high, relativistic effects will be manifested in the elastic differential cross section. Relativistic kinematics are needed for high energy reactions and are easily incorporated into the momentum space-representation of the PW and LS3D equations, where the momentum is simply a number instead of a spatial derivative operator, as in the position space-representation. At relativistic energies, the PW and LS3D models will agree if convergence of the partial wave solution is reached, but both methods should differ from the eikonal results, which are non-relativistic. In the low energy limit, the eikonal method should break down and begin to diverge from the PW and LS3D results, since small angle scattering is not appropriate for such reactions. To examine the effect of kinematics, model results are compared for various nuclear reactions at relativistic and non-relativistic energies.

At relativistic energies, the inner structure of the nucleons may be probed. The multiple scattering theory (MST) upon which the model of interaction is based and the NN transition amplitude do not account for the inner structure of the nucleons. The complications associated with the inner structure of the nucleons are assumed to be included in the parameterizations to experimental NN transition amplitudes.

In this paper, exact formulas of the optical potential in the position space representation are used for light nuclei ($A\le 16$), and the momentum-space representation of the eikonal phase factor is used for heavier nuclei. The PW and LS3D methods are solved with non-relativistic and relativistic kinematics, and comparisons of the models are made for reactions that are relevant to space radiation. Based on the results presented herein, it is recommended that the LS3D method be used for high energy NA reactions and AA reactions at all energies because of its rapid numerical convergence and stability. The effect of the kinematics for projectiles and targets with equal masses and extensive comparisons to experimental data will be communicated in subsequent manuscripts.

The present work is organized as follows. In section II, a theoretical overview of the LS equation, MST, the elastic scattering equation, and the optical potential are reviewed. This is followed by a discussion of the eikonal, PW, and LS3D solution methods in section III.  Comparisons of model results and experimental data are given in section IV. The conclusions are stated in section V.

\section{Theoretical Framework}
The LS equation is an expression for the scattering transition operator
---the fundamental quantity that is used to evaluate the elastic differential, elastic, reaction, 
and total cross sections for nuclear reactions---and is given as  
\begin{equation}
T = V + VG_0^+T, \label{LS}
\end{equation}
where V is the sum of residual two-body interactions for the projectile-target system, and $G_0^+$ is the unperturbed two-body propagator \cite{Maung_Norbury_RMST}.
Using projection operators, Eq. \ref{LS} can be expressed as a coupled system of equations \cite{Feshbach1,Feshbach2}
\begin{align}
T &= U + U P G_0^+P T \label{eq:TU}, \\
U &= V + V Q G_0^+Q U,
\end{align}
where Eq. \ref{eq:TU} is the elastic scattering equation, and $U$ is the optical potential. 
The ground state projector is defined as $P = \ket{\phi_0^{A_P},\phi_0^{A_T}} \bra{\phi_0^{A_P},\phi_0^{A_T}}$, and the excited state projectors are defined $Q = 1 - P$, where 
$\ket{\phi_0^{A_P}}$ is the projectile wave-vector, and $\ket{\phi_0^{A_T}}$ is the target wave-vector.

In the non-relativistic multiple scattering theory (MST), the free Hamiltonian can be separated from the residual interaction, $V$. If the interaction is expressed as the sum of two-body projectile and target nucleon interactions, $v_{ij}$, then the Watson series for the optical potential is given by \cite{Watson, Werneth_RMST},
\begin{equation}
U =  \sum \limits_{i=1}^{A_P} \limits \sum_{j=1}^{A_T} U_{ij},
\end{equation}
with 
\begin{equation}
U_{ij} =  \tilde \tau_{ij} +  \tilde \tau_{ij} QG_0^+ Q\sum_{k \ne i}^{A_P} \limits \sum_{l \ne j}^{A_T}   U_{kl} \label{eq:Uij},
\end{equation}
where $A$ is the number of nucleons in the projectile ($P$) or target ($T$), and 
$\tilde \tau_{ij}$ are the Watson-$\tau$ operators that are expressed as $\tilde \tau_{ij}= v_{ij} + v_{ij} QG_0^+Q \tilde \tau_{ij}$. The Watson-$\tilde \tau$ operators are 
often approximated by the free two-body transition amplitudes (impulse approximation) given by
\begin{equation}
t_{ij} = v_{ij} + v_{ij}g t_{ij}, \label{eq:tijtwobody}
\end{equation}
where $g$ is the free NN Green's function.   
The current work uses the first order (single scattering) approximation for the optical potential, which is given by
\begin{equation}
 U \approx \sum \limits_{i=1}^{A_P} \limits \sum_{j=1}^{A_T} t_{ij}. \label{eq:firstorder}
\end{equation}
Note that even in the first order approximation,  $t_{ij}$ represents an infinite series in terms of $v_{ij}$---which can be seen by iteration of equation \eqref{eq:tijtwobody}---but, in practice, $t_{ij}$ is parameterized to experimental data.

The elastic scattering equation is written
\begin{equation}
T({\bf k'}, {\bf k}) = U({\bf k'}, {\bf k}) + \int  \frac{ U({\bf k'}, {\bf k''})   T({\bf k''}, {\bf k}) }   {E(k) - E(k'') + i \epsilon} d {\bf k''} \label{eq:LS3DFull},
\end{equation}
where $\bf{k}$ $(\bf{k}')$ is the initial (final) momentum in the center of momentum (CM) frame, $E$ is the energy, $k = |\vec k|$ is the relative on-shell momentum, and $i\epsilon$ is imposed to ensure outward scattering boundary conditions. 

The optical potential is found by taking the matrix element of equation \eqref{eq:firstorder},
\begin{equation}
U({\bf k',k)} = \sum \limits_{i=1}^{A_P} \limits \sum_{j=1}^{A_T} \bra{{\bf k'};\phi_0^{A_P} \phi_0^{A_T}} t_{ij} \ket{\phi_0^{A_P} \phi_0^{A_T};{\bf k}}
	      = \xi  \bra{{\bf k'},\phi_0^{A_P} \phi_0^{A_T}} t \ket{\phi_0^{A_P} \phi_0^{A_T},{\bf k}}     \label{eq:uopt1},
\end{equation}
where $\xi = A_P A_T$ using the Watson \cite{Watson} convention.
% or $\xi = xxy$ with the KMT \cite{KMT1959} convention.
Following the work in references \cite{Pickle, Wolfe, Elster}, the optical potential can be expressed as
\begin{equation}
 U({\bf k',k)} = \xi \eta t(e_{\rm NN},q)\rho_P(q)\rho_T(q), \label{eq:optpotfinal}
\end{equation}
where $\vec{q} = \vec{k'} - \vec{k} $, $q=|\vec q|$, $\rho(q)$ is the nuclear matter density, $t(e_{\rm NN},q)$ is the NN transition amplitude,
$e_{\rm NN}$ is the NN CM energy, and $\eta$ is the M\"{o}ller frame transformation factor \cite{Moller,Joachain} used to transform from the AA to NN CM frame.
Nuclear charge densities and the NN transition amplitude are parameterized to experimental data \cite{NASATP2014, Werneth_Finite,NASARP1257,DeVries1,DeVries2,CJP1982,CJP1983}.

The elastic scattering amplitude is related to the transition matrix by \cite{Joachain}
\begin{equation}
f(\theta) = \frac{-(2\pi)^2 \rho}{k} T(k,\theta), % \langle {\bf k'}|T| {\bf k} \rangle,
\end{equation}
where $k = |\vec k|$, $\theta$ is the CM scattering angle, the density of states, $\rho$, is given by
\begin{equation}
\rho = k^2 dk/dE \label{eq:rho},
\end{equation}
and $E$ is the energy. For non-relativistic (NR) kinematics, $E = k^2/2\mu$, where  $\mu = (m_P m_T)/(m_P + m_T)$ is the reduced mass, $m_P$ is the mass of the projectile, and $m_T$ is the mass of the target. When using relativistic (REL) kinematics, $E = \sqrt{k^2 + m_P^2} + \sqrt{k^2 + m_T^2}$.
 Elastic differential cross sections are determined from the scattering amplitude by using
\begin{align}
 \frac{d \sigma}{d\Omega} &= |f(\theta)|^2.
\end{align}

\section{Solution Methods}
The Lippmann Schwinger equation was solved with two approximate methods and a full three-dimensional approach. Approximate solutions include the eikonal method, which employs a forward scattering approximation, and the PW method in the which the transition amplitude is expanded in an infinite series of Legendre polynomials. This section outlines the solution methods and numerical techniques used to solve for the transition matrix and scattering amplitude.

\subsection{Lippmann-Schwinger Partial Wave Solution Method}
The LS equation is often solved with partial wave decomposition, a well-known method that is described in standard texts \cite{Joachain, QMII, Werneth_Finite}. 
In this method, the transition matrix is decomposed into a complete orthonormal set of momenta dependent functions and angular dependent Legendre polynomials. For elastic scattering
\begin{equation}
  T(q)  = \sum_{l=0}^\infty \frac{2l+1}{4 \pi}T_l(k',k) P_l(x), \label{Tsum}
\end{equation}
where $k = |\vec k|$, $k' = |\vec k'|$, $P_l(x)$ are the Legendre Polynomials, $x = \cos(\theta)$, $\theta$ is the angle between $\vec k$ and $\vec k'$, and $q =2k \sin(\theta/2)$.
The angular dependence is integrated, and the solution to the Lippmann-Schwinger equation is found for each partial wave,
\begin{equation}
T_l(k',k) = U_l(k',k) + \int \limits_0^\infty \frac{U_l(k',k'') T_l(k'',k) {k''}^2}{E(k)-E(k'') + i\epsilon} d{k''}, \label{eq:Partial}
\end{equation}
where
\begin{equation}
U_l(k',k) = 2\pi \int \limits_{-1}^{1} U(q) P_l(x) dx. \label{Ul}
\end{equation}

Equation \eqref{eq:Partial} is expressed in terms of its principal value integral, and Gaussian quadrature is used for the momentum integration variable.
Sloan's method \cite{Sloan} is employed for the principal value integral, and the transition amplitude is expressed as a matrix equation for each partial wave, which is solved. The number of partial waves needed for an acceptable tolerance of convergence is not known a priori. Partial waves must be generated until such a tolerance is reached.  

In the current work, the authors use a finite summation formula for the transition amplitude, which is given by \cite{Werneth_Finite}
\begin{equation}
T(q) = \sum \limits_{l=0}^{l_{\rm max}} \frac{2l + 1}{4 \pi}\left[ T_l(k,k) - U_l(k,k)\right] + U(q),
\end{equation}
where $l_{\rm max}$ represents a finite angular momentum that is reached when $T_l(k,k) \approx U_l(k,k)$ according to a pre-defined tolerance of $|T_l -U_l| \le 10^{-4}$ \%.

\subsection{Lippmann-Schwinger 3D Solution Method}
The three-dimensional Lippmann-Schwinger (LS3D) solution method avoids the numerical difficulties associated with the PW method and has been used for relatively low energy reactions \cite{LS3D_1,LS3D_2,LS3D_3,LS3D_4,LS3D_3body}. This section outlines the LS3D equation and the solution methods.

 If one considers only central potentials in equation \eqref{eq:LS3DFull}, then both $T$ and $V$ are scalar functions; that is,  $f({\bf k'}, {\bf k}) = f(k',k, \hat k' \cdot \hat k)$ for some function $f$, where $\hat k $ $(\hat k')$ represents the unit vector associated with $\vec k$ $(\vec k')$.
The possible scalar products of the LS equation are as follows \cite{LS3D_1,LS3D_3}:
\begin{eqnarray}
x' &\equiv& \hat k' \cdot \hat k \\
x'' &\equiv& \hat k'' \cdot \hat k \nonumber \\ 
y &\equiv&  \hat k'' \cdot \hat k'. \nonumber
\end{eqnarray}

\noindent
The incoming momentum, ${\bf k}$, is taken to be in the direction of the $z$-axis, and the azimuthal angle between ${\bf k}$ and ${\bf k'}$ is set to zero: $\phi' = 0$; therefore, $y$ may be expressed as a function of $x'$, $x''$, and $\phi''$ \cite{LS3D_1,LS3D_3},
\begin{equation}
y = x' x'' + \sqrt{1-{x'}^2} \sqrt{1-{x''}^2} \cos \phi'',
\end{equation}
and the LS3D equation is given by \cite{LS3D_1}
\begin{align}
T(k',k,x') &= U(k',k,x') \label{eq:finalLS3D} \\ 
&~~~~+\int \limits_0^\infty {k''}^2 d {k''} \int \limits_{-1}^{1} d{x''} \frac{ \bar U(k',x',k'',x'')  T(k'',k,x'')}{E(k)-E(k'') + i\epsilon} \nonumber,
\end{align}
where \cite{LS3D_1}
\begin{equation}
	\bar U(k',x',k'',x'') \equiv \int \limits_0^{2 \pi} U(k',k'',y) d\phi''.
\end{equation}

The numerical implementation of the LS3D method proceeds in the same manner as the PW method, 
but there are now two additional integration variables over azimuthal and polar angles. 
The azimuthal dependence only occurs in the potential and is integrated with 40 Gaussian quadrature points. 
As was seen with the PW method, the principal value integral over momenta is handled with Sloan's method \cite{Sloan}, and the transition amplitude is expressed as a matrix equation, which is solved. The solution, $T(k'',x'')$, corresponds to the transformed Gaussian quadrature points associated with the integral. These results are substituted back into equation \eqref{eq:finalLS3D} to obtain the transition amplitude at the specified final momentum ($k'$) and angle ($x'$).

It has been observed that the transition amplitude for reactions with large on-shell 
momenta---including high energy NA reactions and AA reactions at every energy---do not converge 
efficiently if the integration ranges of both momenta and polar angles are not restricted to regions 
that give significant contributions to the LS equation. 
The momenta which give non-zero contributions are estimated from the range of the optical potential 
and tend to be localized near the on-shell momentum, $k$. For numerical efficiency and to ensure convergence,
the integrations are truncated accordingly.  The number of Gaussian quadrature points for the LS3D solution method was increased to a maximum of 44 points such that the total elastic cross sections changed less than 1\% for all reactions with energies up to 100 GeV/n.

\subsection{Eikonal Solution Method}
The eikonal approximation is used for high energy, small angle scattering to calculate elastic, reaction, total, and elastic differential cross sections \cite{Joachain, Glauber}. To compute cross sections with the eikonal method, one solves for the
 eikonal scattering amplitude, $f(\theta)$, which is given as \cite{Joachain}
\begin{equation}
  f(\theta) = \frac{k}{i} \int \limits_0^\infty J_0(2k\sin(\theta/2)) \left[ e^{i\chi(k,b)} -1\right] b ~db, \label{eq:ftheta}
\end{equation}
where $k$ is the relative momentum of the projectile-target system in the CM frame, $J_0$ is the ordinary cylindrical Bessel function, $\theta$ is the scattering angle in the CM frame, $b$ is the impact parameter, and $\chi(k,b)$ is the eikonal phase shift function, the latter of which is obtained by integrating over the optical potential, $U(b,z)$ \cite{Joachain}:
\begin{equation}
\chi(k,b) = -\frac{1}{2k} \int \limits_{-\infty}^{\infty} U(b,z) dz. \label{eq:chi}
\end{equation}  
The $z$-integration is taken to be in the same direction as the initial wave vector of the incident projectile.
The optical potential in equation \eqref{eq:chi} is the Fourier transform of equation \eqref{eq:optpotfinal} \cite{NASARP1257,Townsend}

The numerical evaluation of equation \eqref{eq:chi} is inefficient when the six-dimensional position-space integral of the optical potential is solved. In the present work, formulas of the optical potential are used for light nuclei ($A\le 16$) \cite{NASATP2014}, and the eikonal phase function is written in the momentum-space representation for heavier nuclei \cite{NASATP2014},
\begin{equation}
\chi(k,b) = -\frac{\pi}{k}\int \limits_0^\infty dq \int \limits_0^{2 \pi}  q \, U(|{\bf q}|) e^{-iqb \cos \phi} d \phi. \label{eq:chifinal}
\end{equation}
The advantage of equation (\ref{eq:chifinal}) is that the optical potential is in the momentum-space representation, 
and the $z$-integration need not be performed. 
Instead, the 7-dimensional integral for $\chi$ has been reduced to 2-dimensions over the magnitude of the momentum transfer, 
$q$, and the angle, $\phi$, between the momentum transfer and the impact parameter. 
This result significantly increases the efficiency for the numerical evaluation of $\chi$.

Although the momentum space method for the eikonal phase function is much more efficient than the position space calculation, 
additional interpolation over the impact parameter and momentum transfer was performed for additional numerical efficiency. 
Convergence of the total elastic cross sections was used to establish the number of Gaussian quadrature points used for integration.
The number of Gaussian quadrature points for the eikonal solution method was increased to a maximum of 100 points such that the total elastic cross sections
changed less than 1\% for all reactions with energies up to 100 GeV/n.

\section{Results}
In the results that follow, each model uses the same set of fundamental parameterizations for the nuclear matter densities and the NN transition amplitude. Harmonic well and two-parameter Fermi (Wood-Saxon) nuclear charge data are taken from references \cite{DeVries1,DeVries2} and are normalized to matter densities as described in reference \cite{NASARP1257}. When data are not available for the two-parameter Fermi densities, a nuclear droplet model \cite{Myers} is used for parameter estimates. Nuclei are assumed to be near the beta stability curve. 
The NN transition amplitude used in the current work is described in reference \cite{NASATP2014} and depends on parameterizations of the NN cross sections, slope parameter, and real to imaginary ratio of the transition amplitude. The NN cross sections are taken from reference \cite{NASATP}, and the slope parameter is from reference \cite{NASATP2014}.

\begin{figure*}
\begin{center}
\subfigure{\includegraphics[scale=0.33]{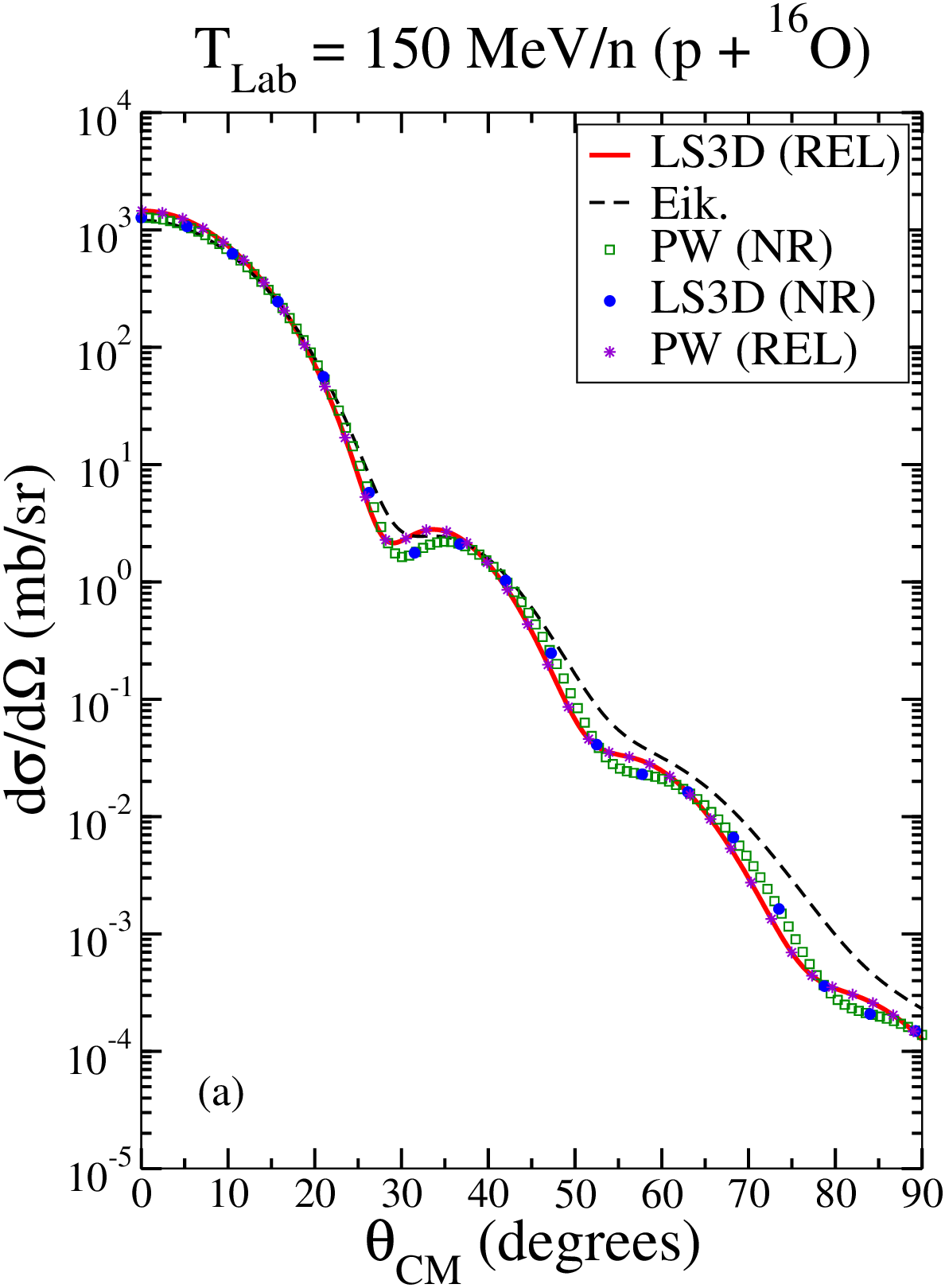}} \hspace{0.5cm}
\subfigure{\includegraphics[scale=0.33]{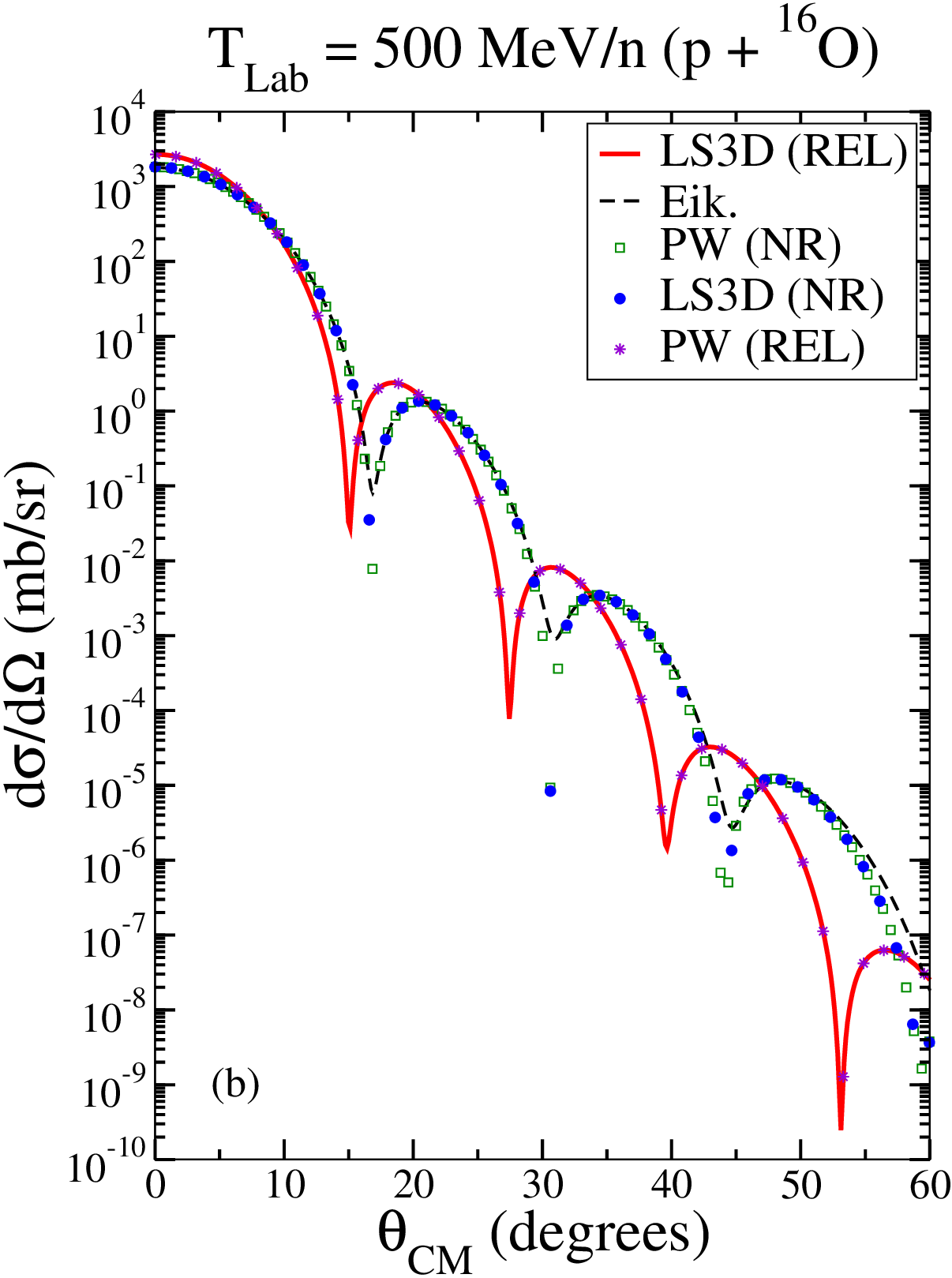}}  \\
\subfigure{\includegraphics[scale=0.33]{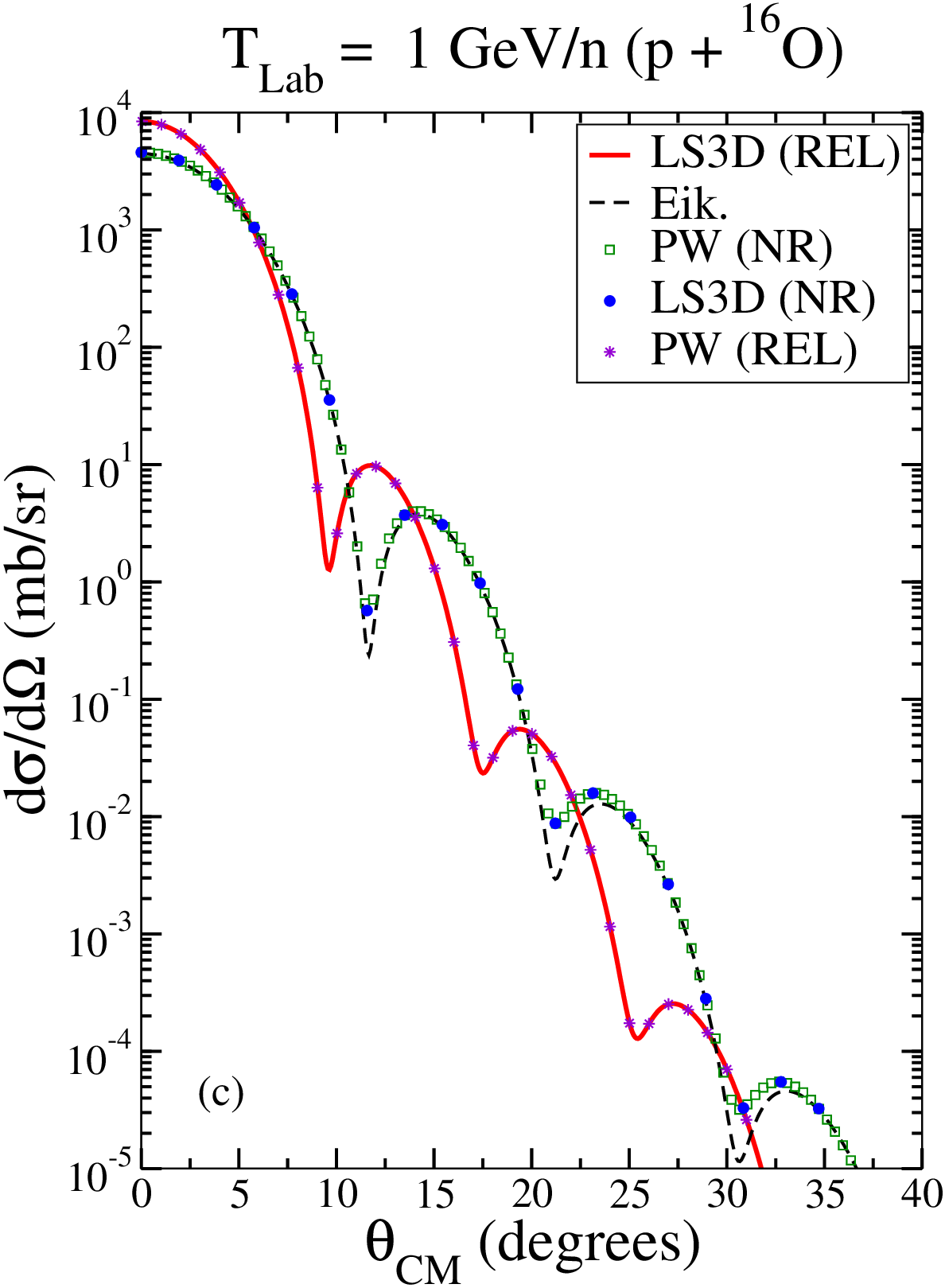}} \hspace{0.5cm}
\subfigure{\includegraphics[scale=0.33]{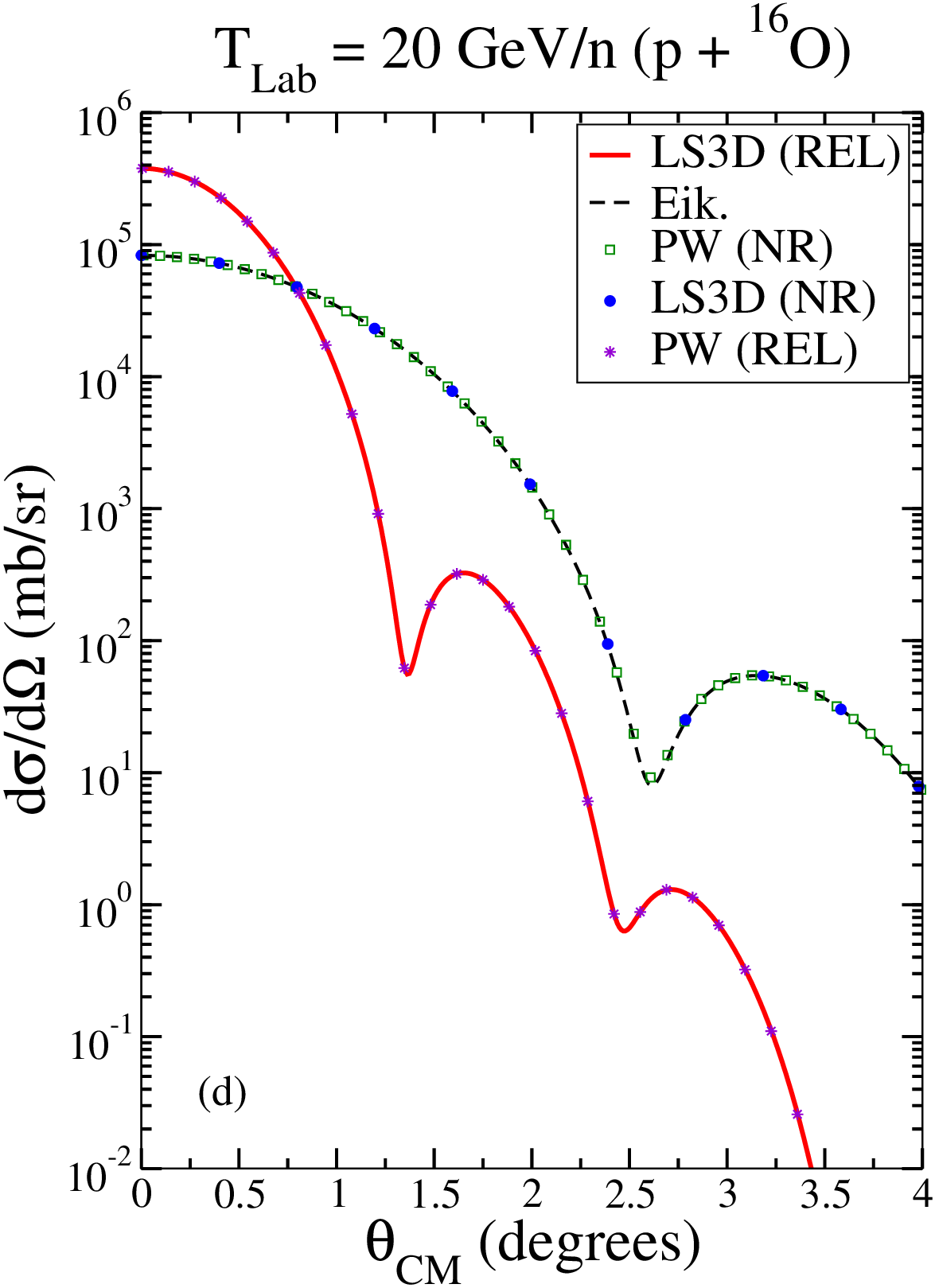}} 
\caption[Elastic differential cross sections for p + $^{16}$O reactions]{Elastic differential cross sections for p + $^{16}$O reactions for projectile lab kinetic energies of (a) 150 MeV (b) 500 MeV (c) 1000 MeV and (d) 20000 MeV. 
Eik. represents eikonal, LS3D represents three-dimensional Lippmann-Schwinger, and PW represents partial wave. Non-relativistic results are denoted (NR) and relativistic results are denoted (REL). 
}
\label{pOfig}
\end{center}
\end{figure*}

\begin{figure*}
\begin{center}
\subfigure{\includegraphics[scale=0.33]{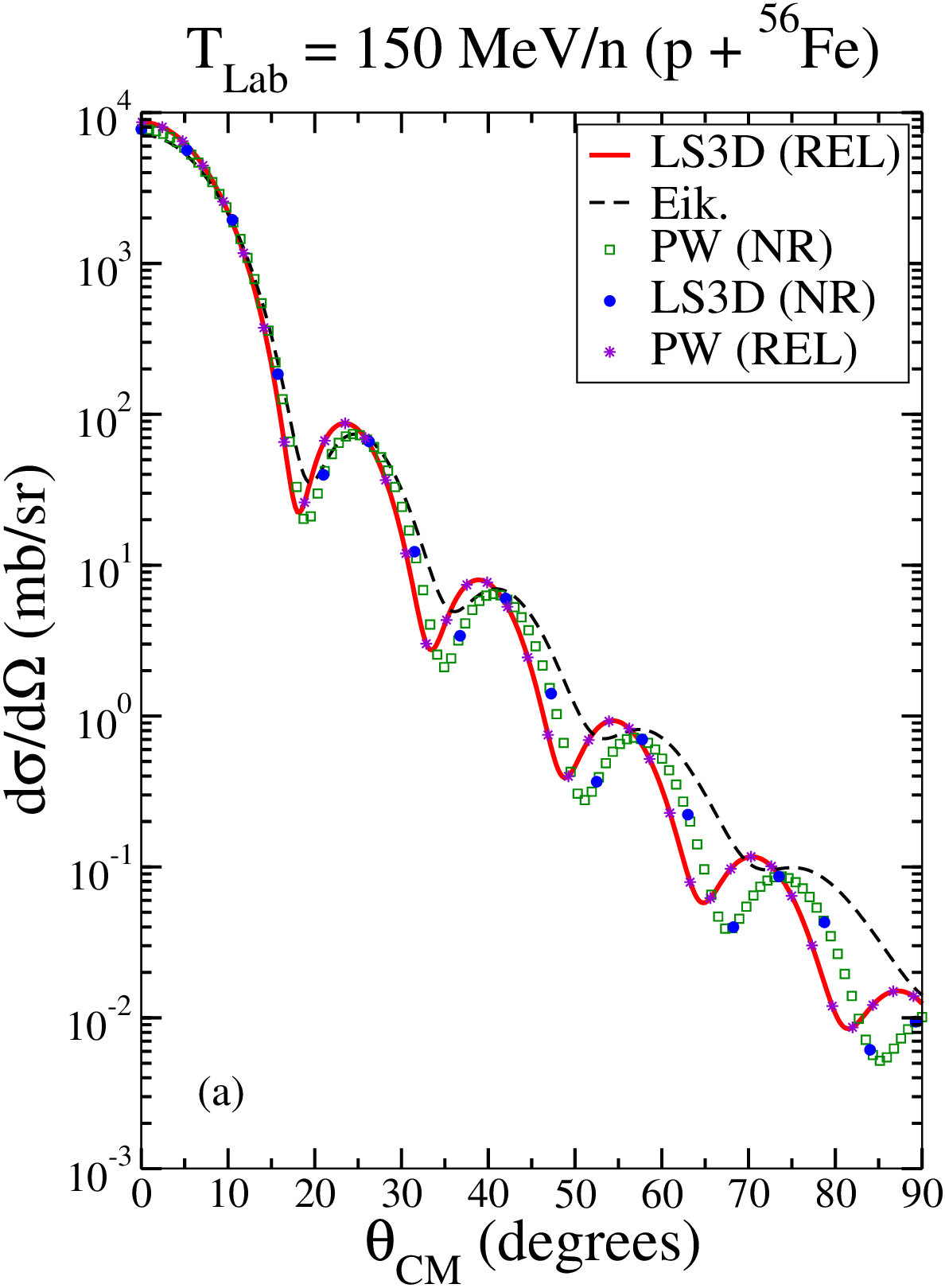}} \hspace{0.5cm}
\subfigure{\includegraphics[scale=0.33]{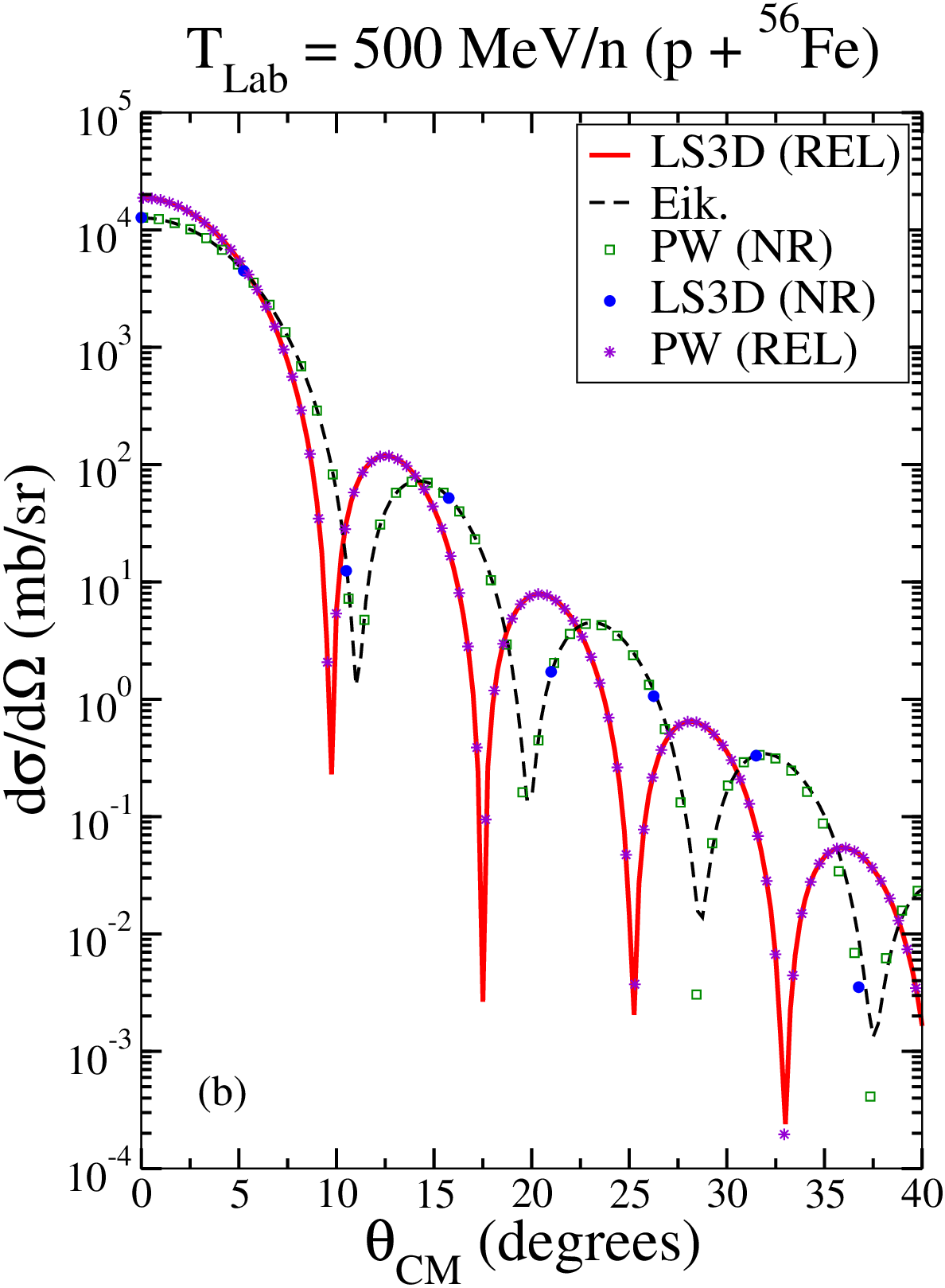}}  \\
\subfigure{\includegraphics[scale=0.33]{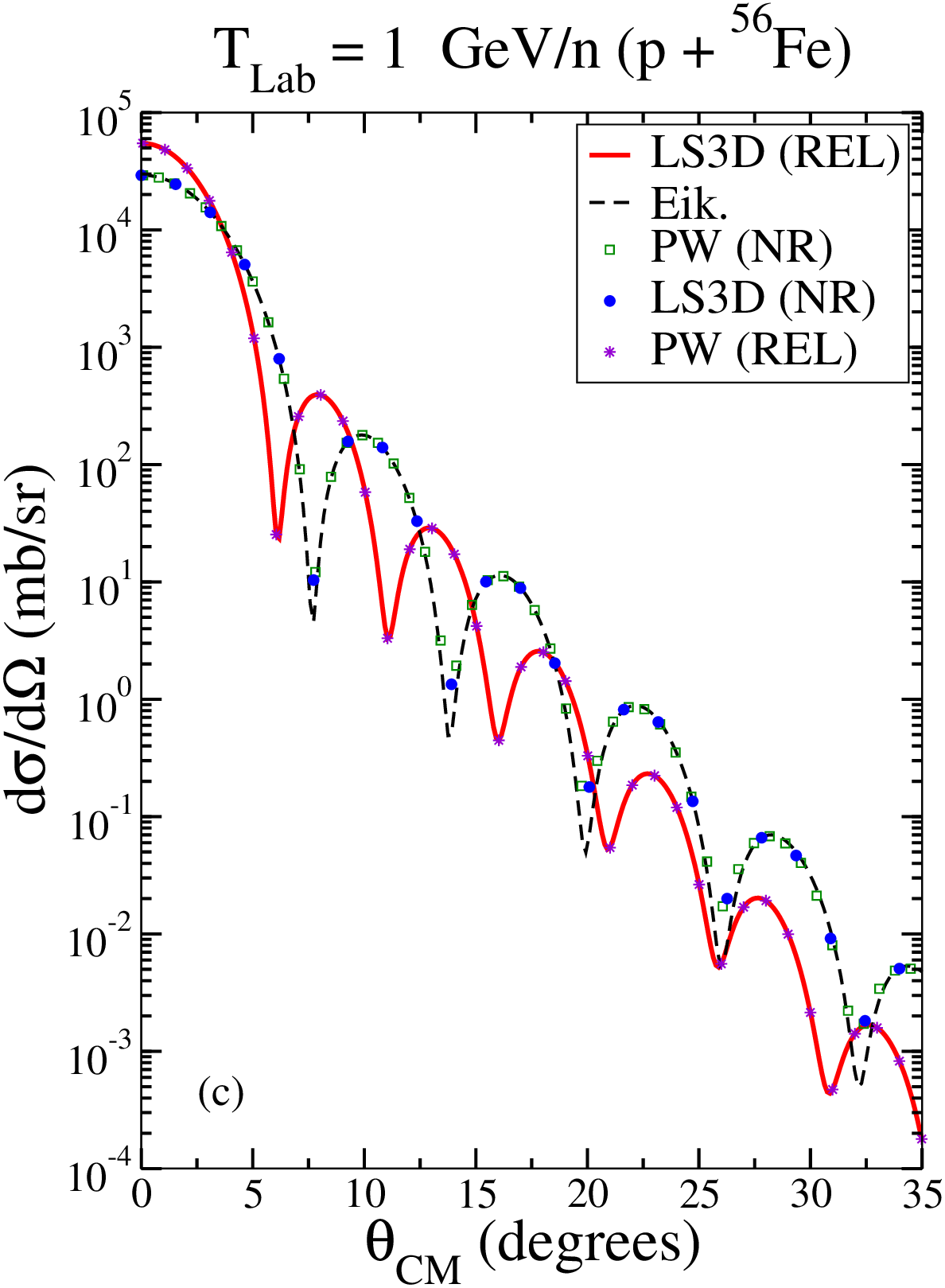}} \hspace{0.5cm}
\subfigure{\includegraphics[scale=0.33]{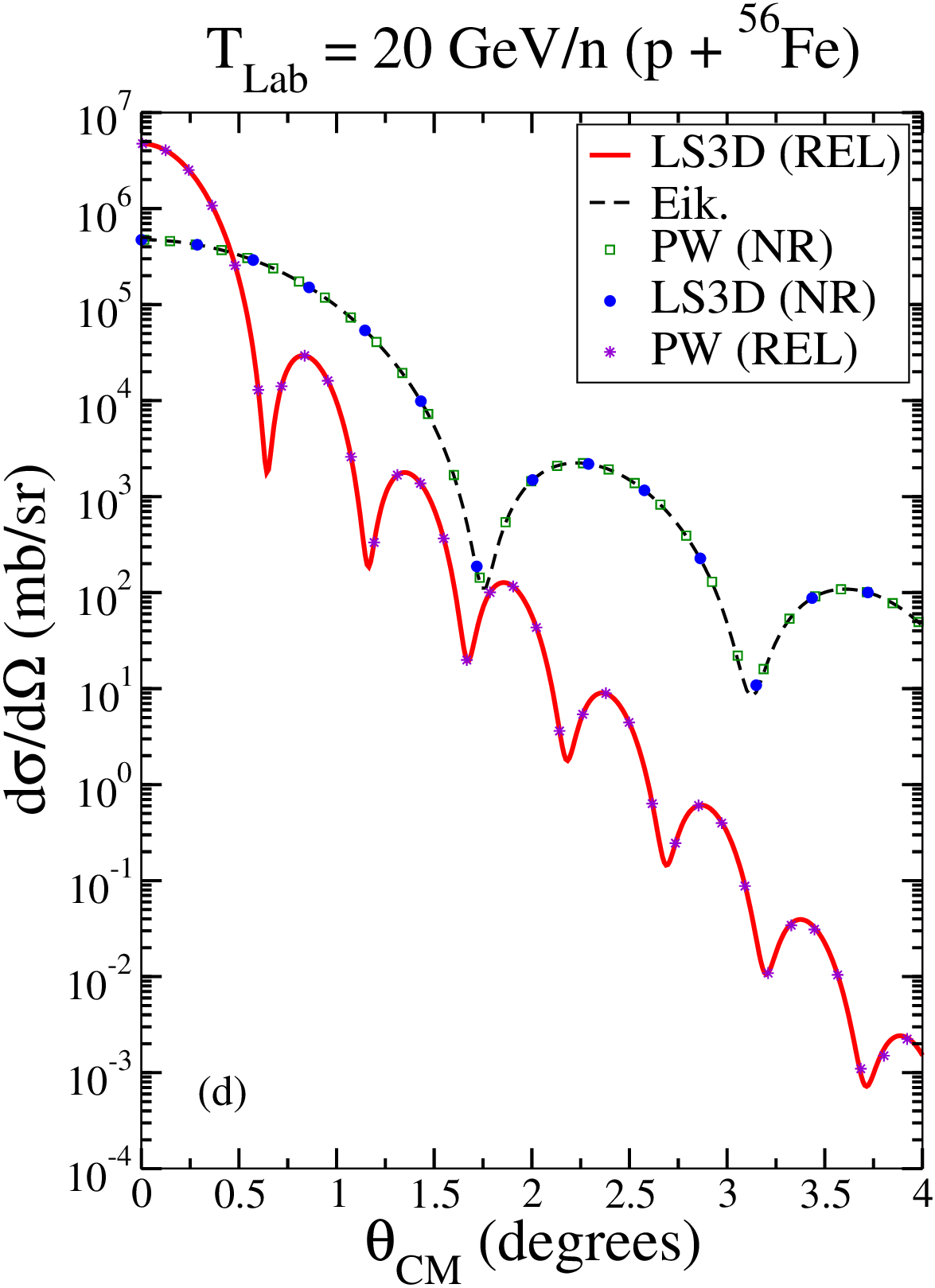}} 
\caption[Elastic differential cross sections for p + $^{56}$Fe reactions]{Elastic differential cross sections for p + $^{56}$Fe reactions for projectile lab kinetic energies of (a) 150 MeV (b) 500 MeV (c) 1000 MeV and (d) 20000 MeV. Eik. represents eikonal, LS3D represents three-dimensional Lippmann-Schwinger, and PW represents partial wave. Non-relativistic results are denoted (NR) and relativistic results are denoted (REL).
}
\label{pFefig}
\end{center}
\end{figure*}

\begin{figure*}
\begin{center}
\subfigure{\includegraphics[scale=0.33]{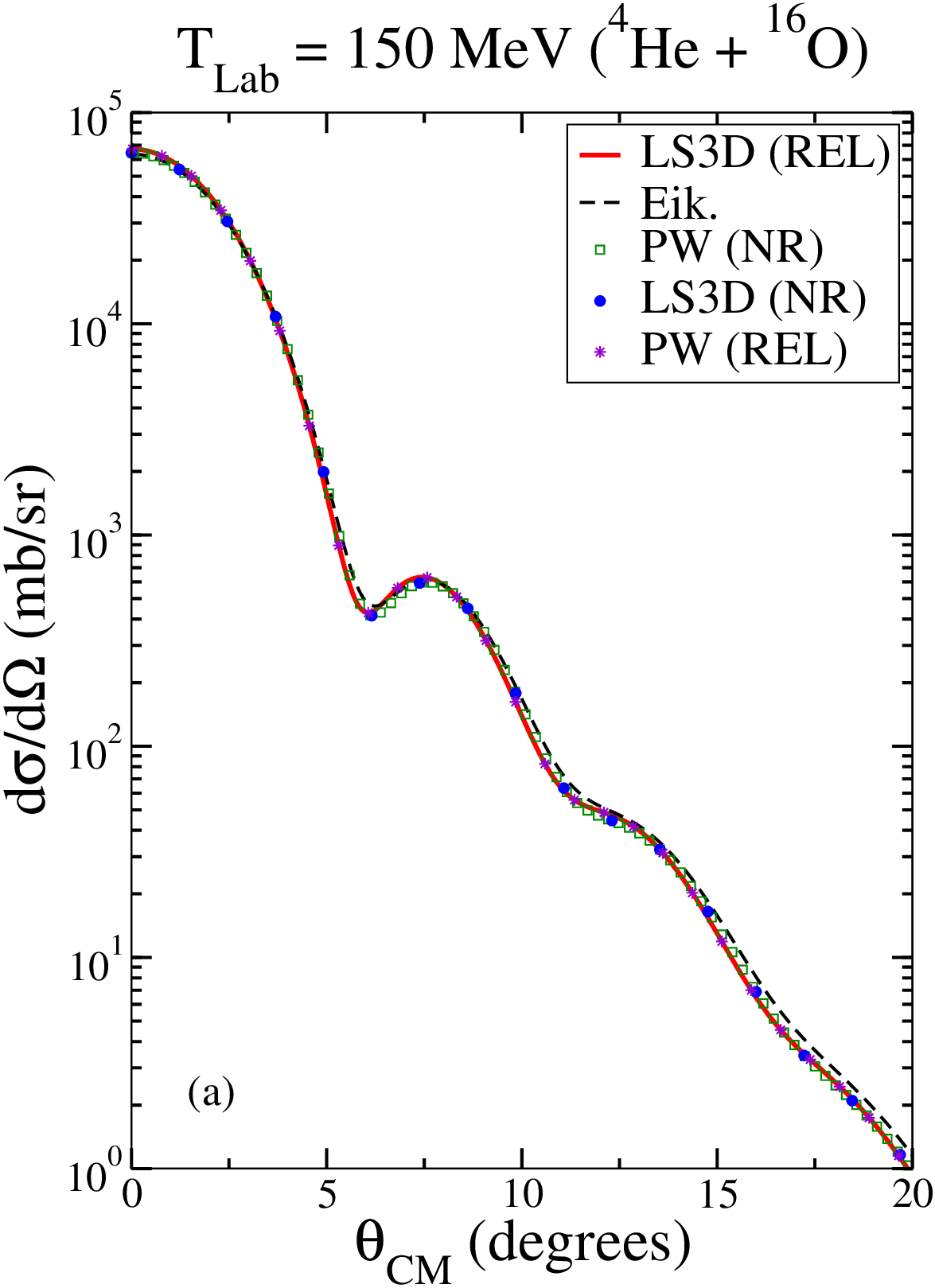}} \hspace{0.5cm}
\subfigure{\includegraphics[scale=0.33]{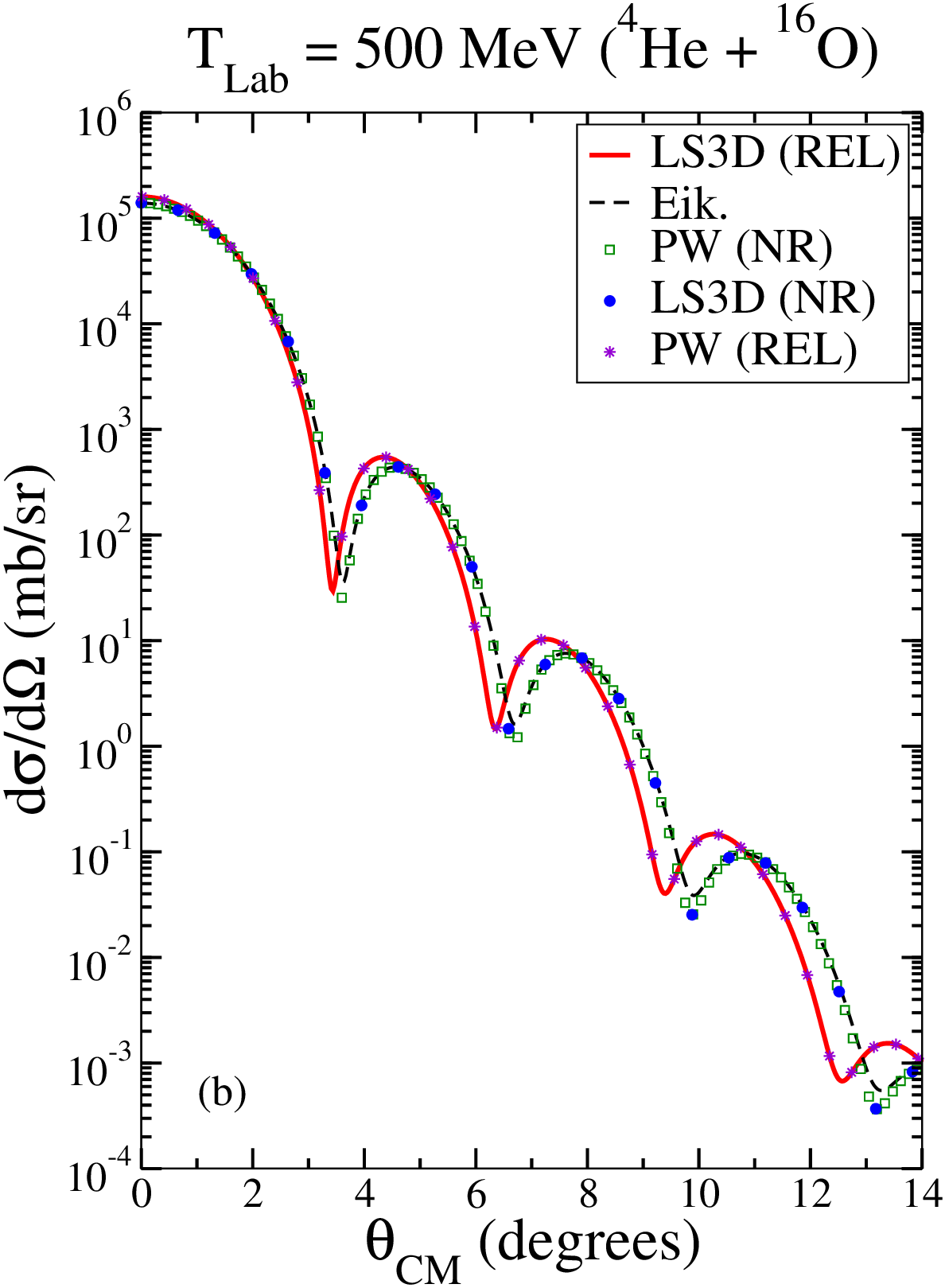}}  \\
\subfigure{\includegraphics[scale=0.33]{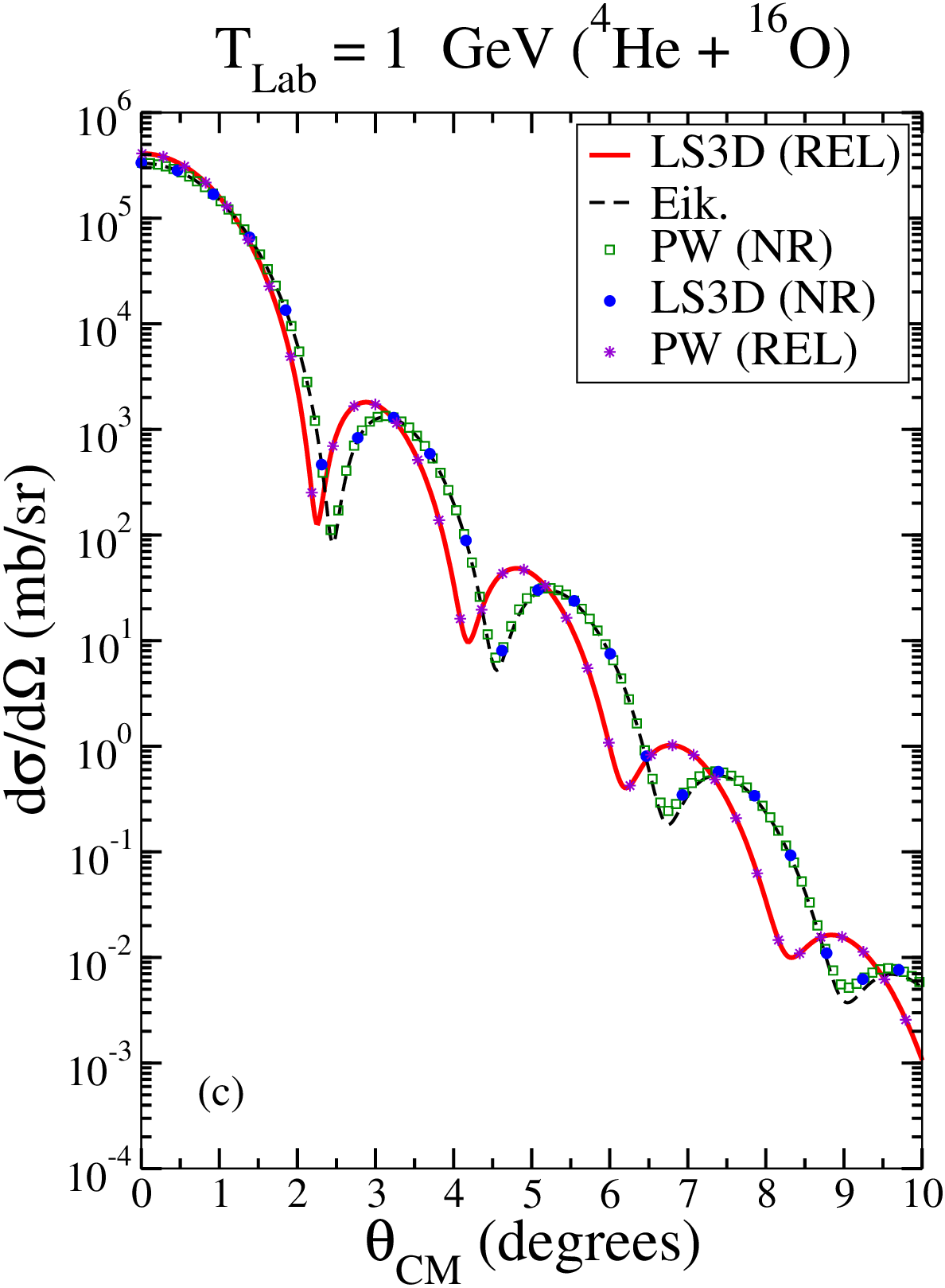}} \hspace{0.5cm}
\subfigure{\includegraphics[scale=0.33]{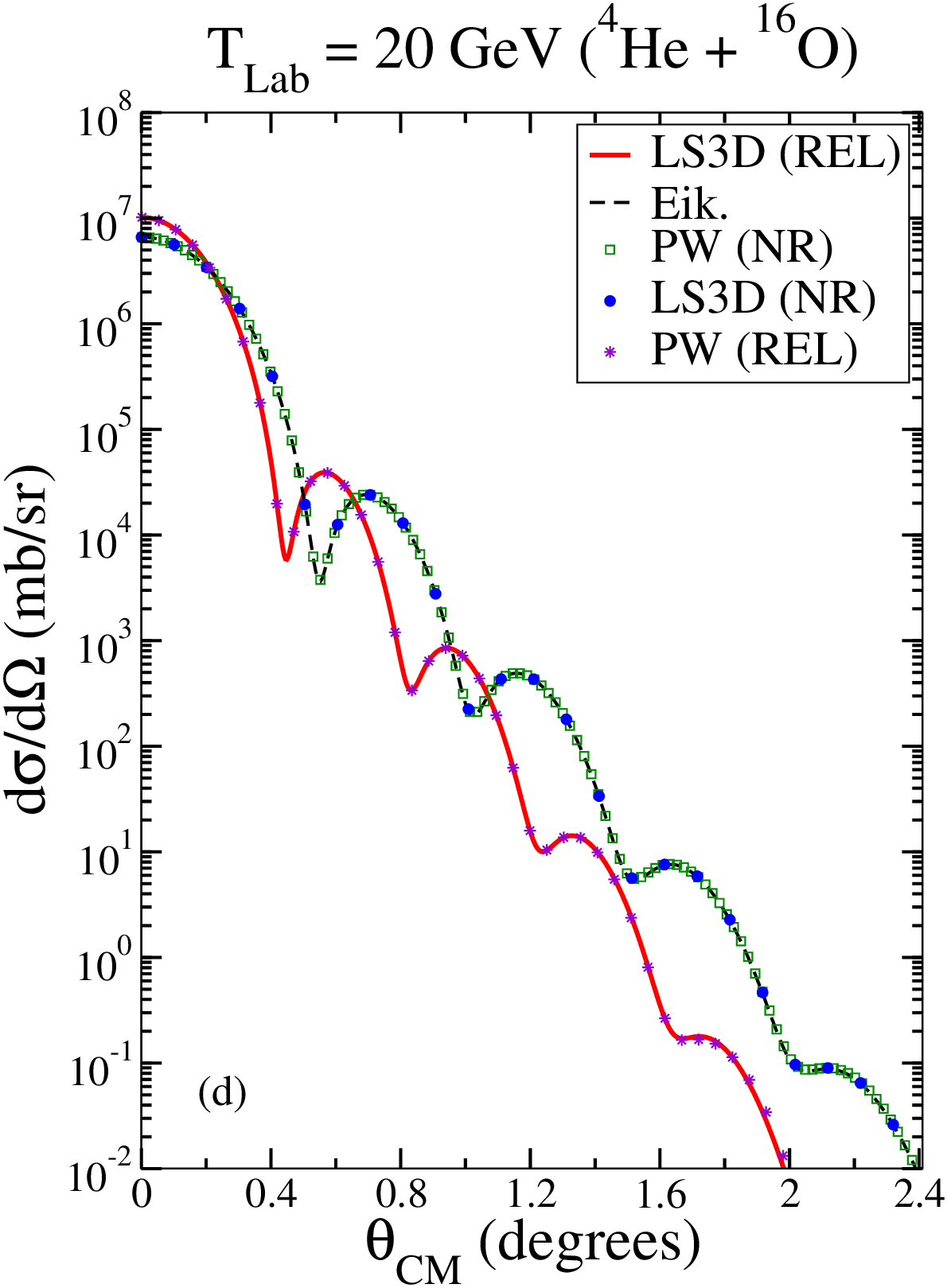}} 
\caption[Elastic differential cross sections for $^{4}$He + $^{16}$O reactions]{Elastic differential cross sections for $^{4}$He + $^{16}$O reactions for projectile lab kinetic energies of (a) 150 MeV/n (b) 500 MeV/n (c) 1000 MeV/n and (d) 20000 MeV/n.
Eik. represents eikonal, LS3D represents three-dimensional Lippmann-Schwinger, and PW represents partial wave. Non-relativistic results are denoted (NR) and relativistic results are denoted (REL).
}
\label{HeOfig}
\end{center}
\end{figure*}

\begin{figure*}
\begin{center}
\subfigure{\includegraphics[scale=0.33]{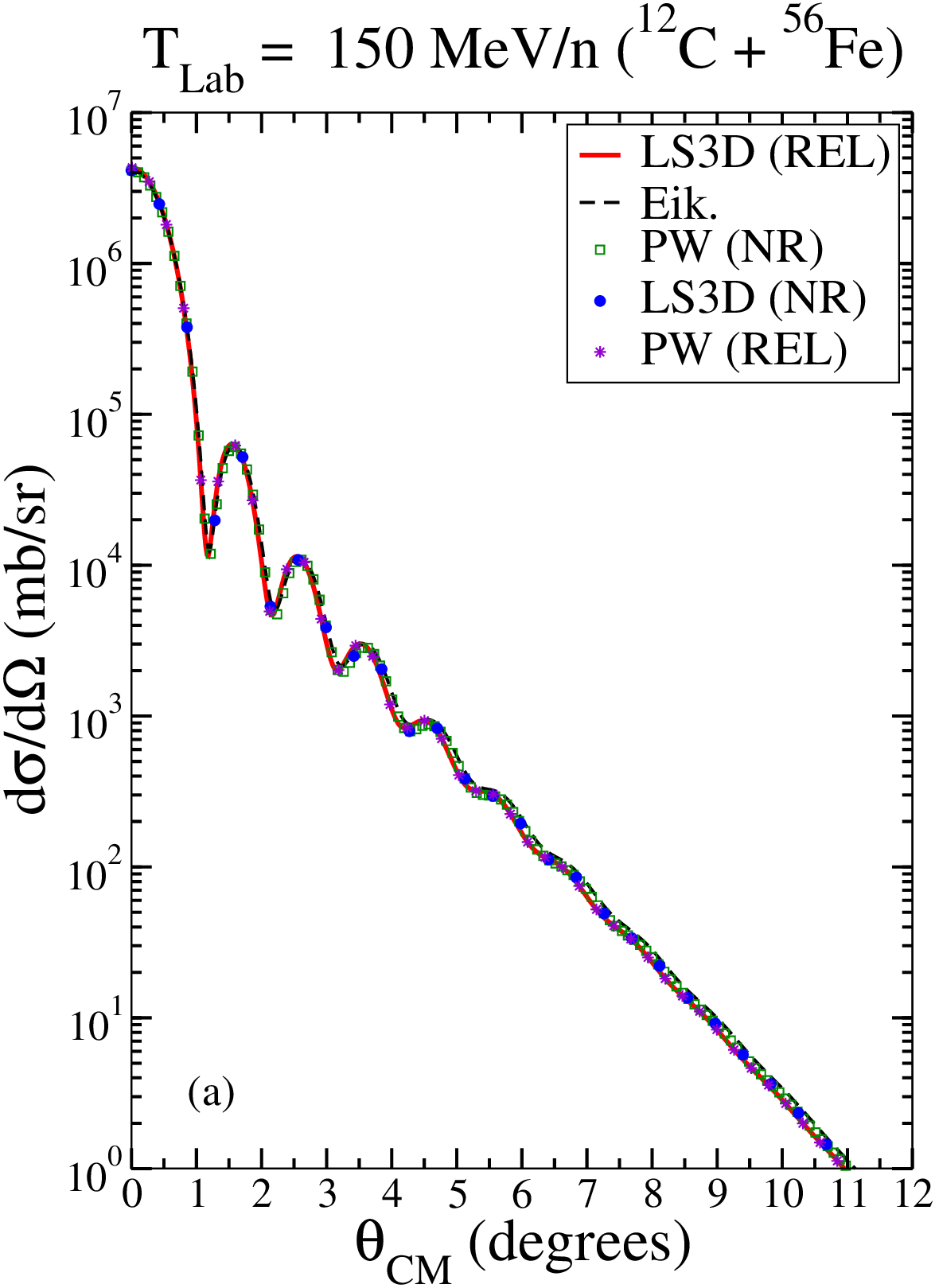}} \hspace{0.5cm}
\subfigure{\includegraphics[scale=0.33]{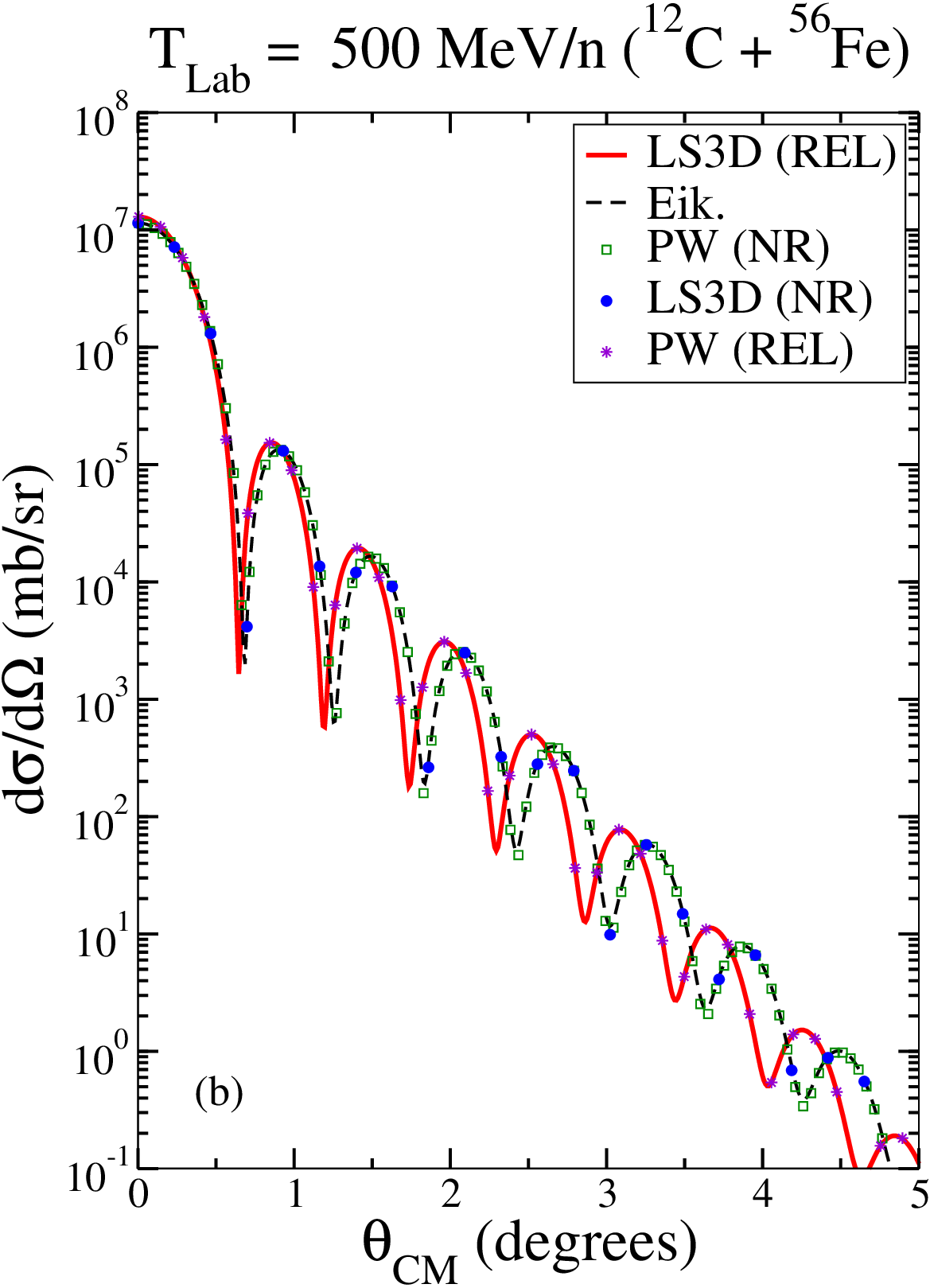}}  \\
\subfigure{\includegraphics[scale=0.33]{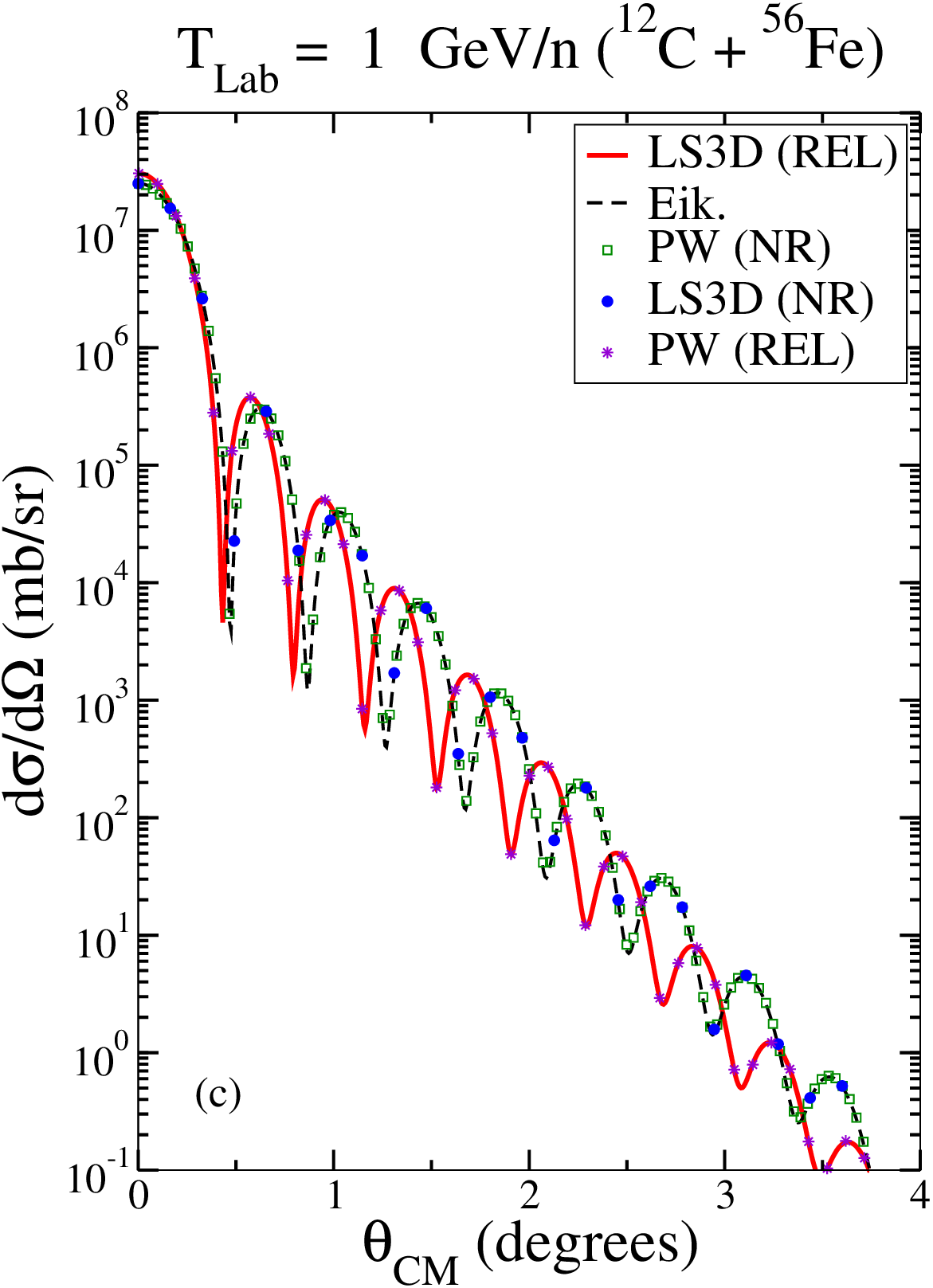}} \hspace{0.5cm}
\subfigure{\includegraphics[scale=0.33]{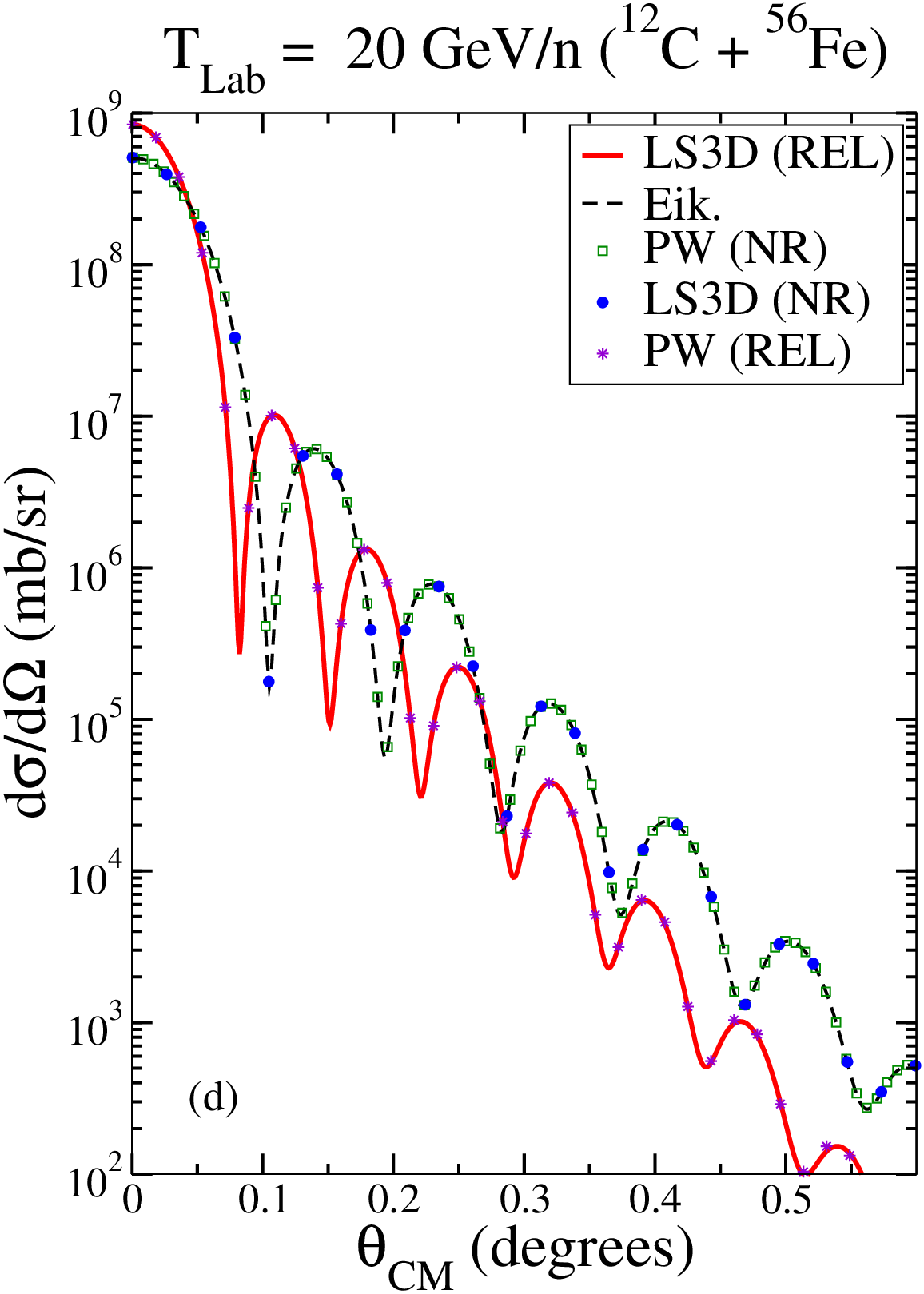}} 
\caption[Elastic differential cross sections for $^{12}$C + $^{56}$Fe reactions]{Elastic differential cross sections for $^{12}$C + $^{56}$Fe reactions for projectile lab kinetic energies of (a) 150 MeV/n (b) 500 MeV/n (c) 1000 MeV/n and (d) 20000 MeV/n.
Eik. represents eikonal, LS3D represents three-dimensional Lippmann-Schwinger, and PW represents partial wave. Non-relativistic results are denoted (NR) and relativistic results are denoted (REL).
}
\label{CFefig}
\end{center}
\end{figure*}

In Figs. \ref{pOfig}-\ref{CFefig}, NA and AA elastic differential cross sections are shown at energies that are relevant to space radiation applications, including p + $^{16}$O, p + $^{56}$Fe, $^4$He + $^{16}$O, and $^{12}$C + $^{56}$Fe reactions at lab projectile kinetic energies of 150, 500, 1000, and 20,000 MeV/n. Results are indicated non-relativistic by (NR) and relativistic by (REL).  LS3D (REL) results are given as a solid red line; a dashed, black line is used for eikonal results, denoted (Eik); a green square represents the PW (NR) results; a solid blue circle indicates LS3D (NR) results; and a violet asterisk is for PW (REL) results. Note that the Coulomb interaction has not been included in this analysis.

Excellent agreement between PW and LS3D results are seen in Figs. \ref{pOfig}-\ref{CFefig} for each kinematic selection for energies greater than 150 MeV. The p + $^{16}$O and p + $^{56}$Fe reactions at 150 MeV/n in Figs. \ref{pOfig} and \ref{pFefig} show slight disagreements between NR PW and LS3D codes and eikonal results. This is likely the result of the forward scattering approximation used in the eikonal method, since very light projectiles may deviate from forward scattering at low energy. The slight disagreement between the eikonal and NR PW and LS3D codes is not observed for the heavier nuclei in Figs. \ref{HeOfig} and \ref{CFefig}, where the small angle scattering approximation is more appropriate.

The next obvious feature is that of the relativistic shift observed in Figs. \ref{pOfig}-\ref{CFefig}. The magnitude of the differential cross section is larger at smaller angles as compared to the NR cases. The effect is more pronounced at higher energies, as expected, but is also driven by projectile and target mass differences. A comparison of Figs. \ref{pOfig} and \ref{HeOfig} shows that the relativistic effect is more pronounced for the p + $^{16}$O reaction, which has larger mass difference than the $^4$He + $^{16}$O
system. Ultimately, the relativistic effects can be tracked back to kinematic differences in the relative on-shell momentum. 

\begin{figure*}
\begin{center}
\subfigure{\includegraphics[scale=0.33]{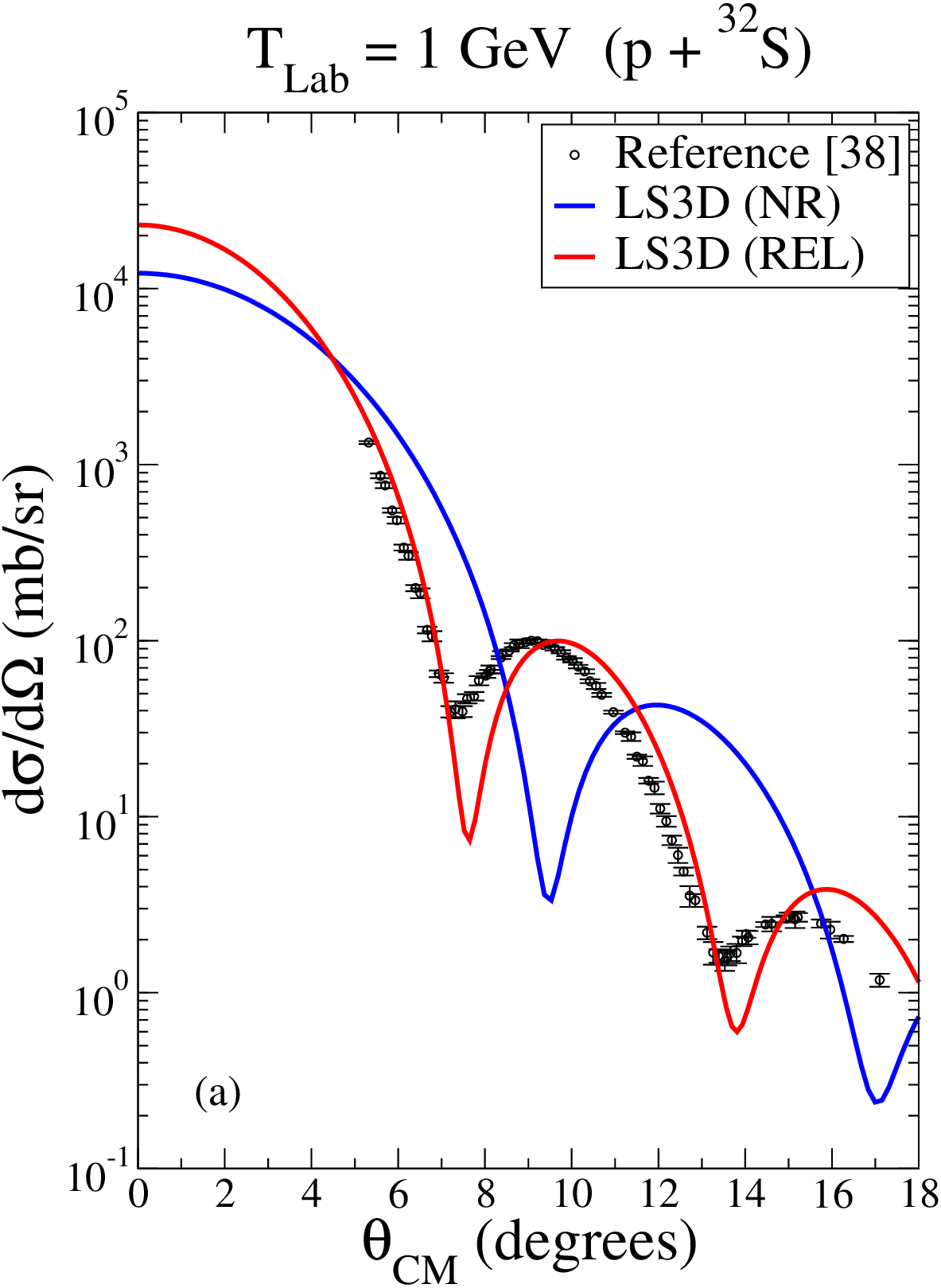}} \hspace{0.5cm}
\subfigure{\includegraphics[scale=0.33]{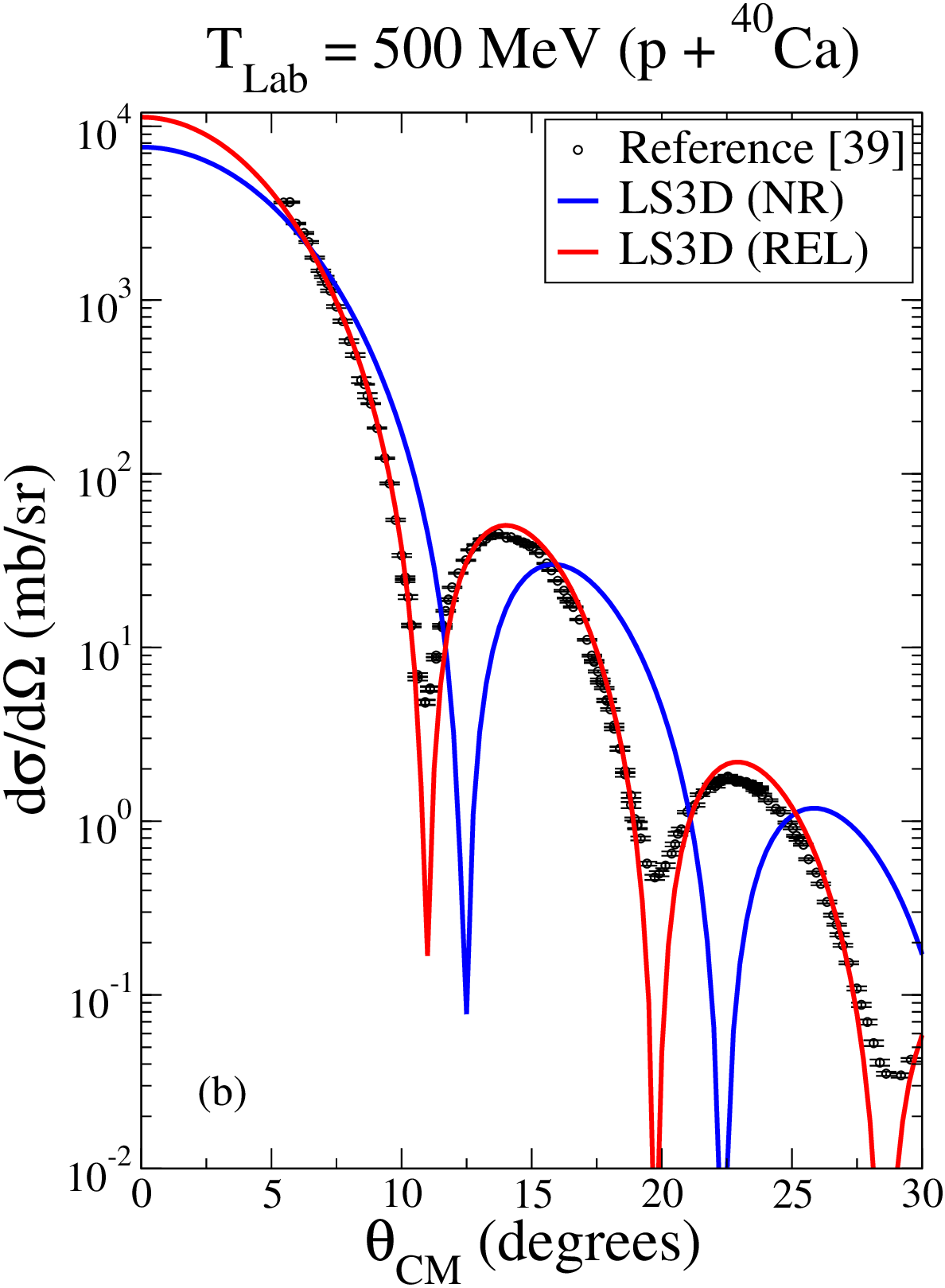}}  \\
\subfigure{\includegraphics[scale=0.33]{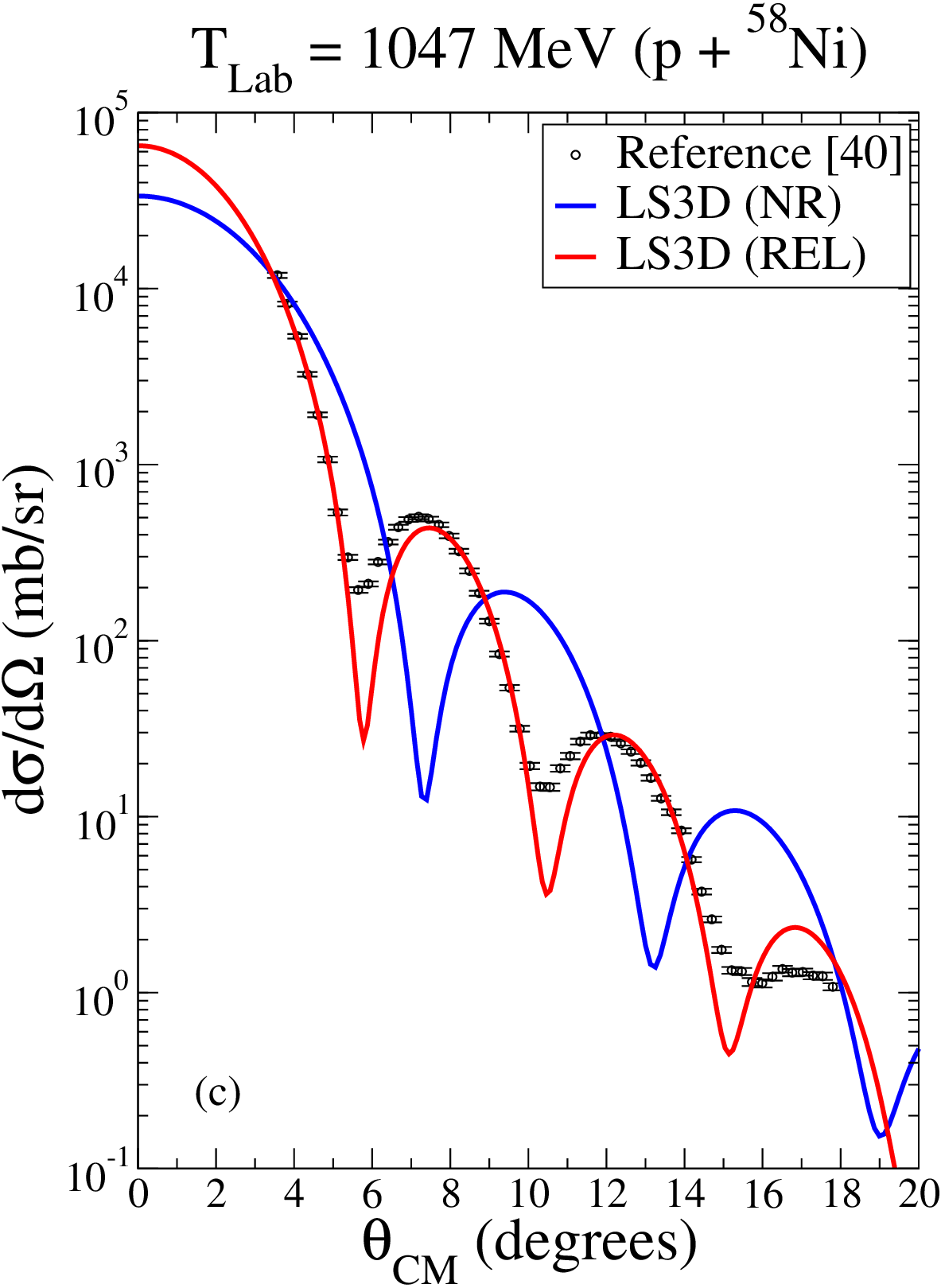}} \hspace{0.5cm}
\subfigure{\includegraphics[scale=0.33]{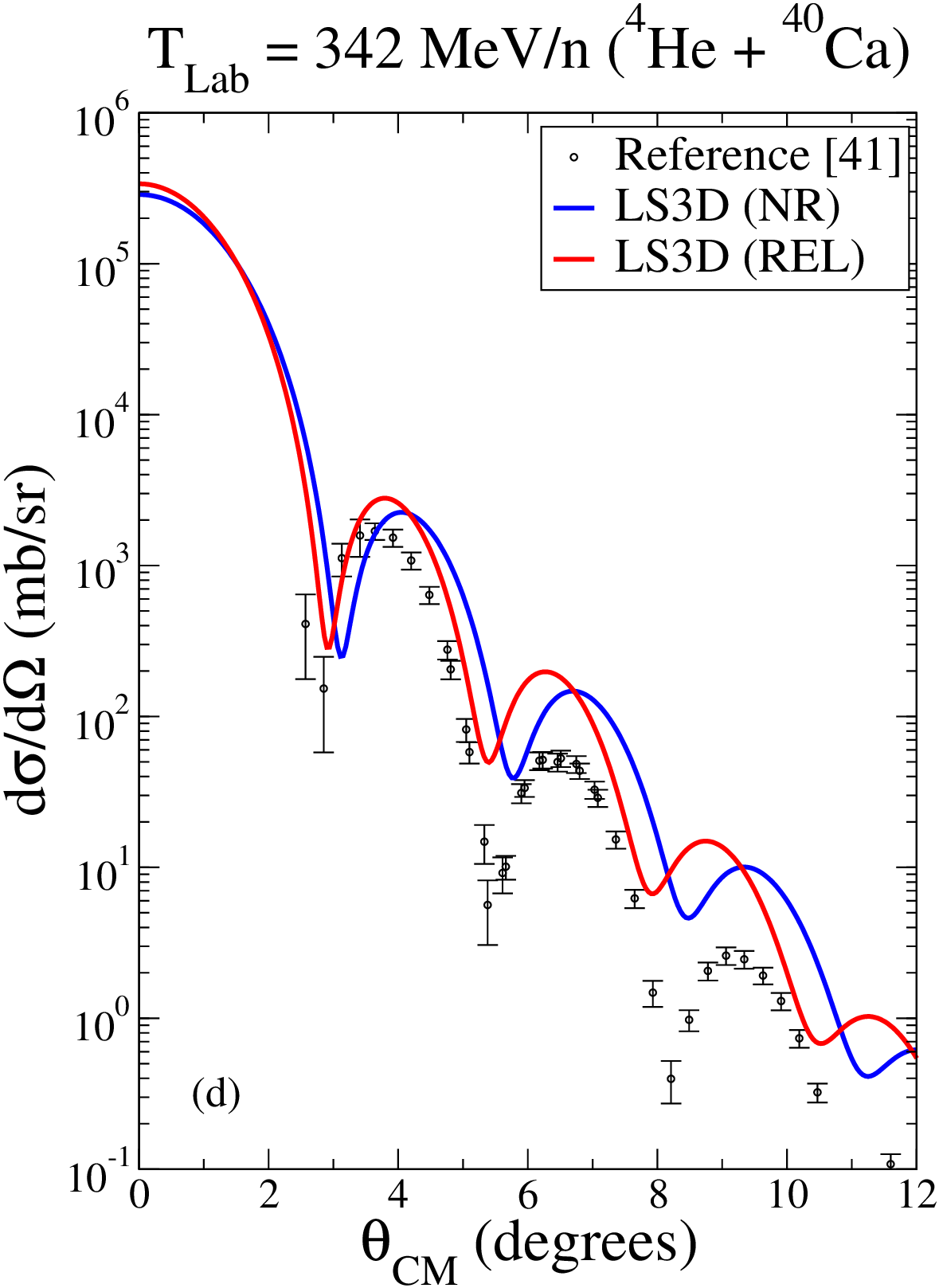}} 
\caption[Elastic Differential Cross Section Comparisons]{Elastic differential cross sections for (a) p + $^{32}$S at $T_{\rm Lab} = 1$ GeV \cite{pS32} (b) p + $^{40}$Ca at $T_{\rm Lab} = 500$ MeV \cite{pCa40} (c) p + $^{58}$Ni at $T_{\rm Lab} = 1$ GeV \cite{pNi58} and (d) $^{4}$He + $^{40}$Ca at $T_{\rm Lab} = 347$ MeV/n \cite{HeCa40}.}
\label{expfig}
\end{center}
\end{figure*}

As an example of the LS3D method and illustration of the relativistic shift, comparisons to experimental data \cite{pS32,pCa40,pNi58, HeCa40} are performed. Fig. \ref{expfig} shows the elastic differential cross sections of the following reactions:  (a) p + $^{32}$S at $T_{\rm Lab} = 1$ GeV \cite{pS32} (b) p + $^{40}$Ca at $T_{\rm Lab} = 500$ MeV \cite{pCa40} (c) p + $^{58}$Ni at $T_{\rm Lab} = 1$ GeV \cite{pNi58} and (d) $^{4}$He + $^{40}$Ca at $T_{\rm Lab} = 347$ MeV/n \cite{HeCa40}. NR results are indicated with a solid red line, and REL results are indicated with a solid blue line. In each case, there is better agreement with experiment when relativistic kinematics are used. Since the fundamental parameterizations are based on small-angle scattering data, the results are in better agreement with the measured differential cross section data at forward scattering angles. Also note that spin-dependence and medium effects have not been included, which may account for differences between the model and the experimental data in Fig. \ref{expfig} (d).

\section{Conclusions}
The eikonal, PW, and LS3D methods have been compared for NA and AA reactions for reactions relevant for space radiation applications.  Numerical convergence of the eikonal method is readily achieved when formulas of the optical potential are used for light nuclei ($A\le16$) and the momentum-space representation of the optical potential is used for heavier nuclei \cite{NASATP2014}. The LS formalism has an advantage over the eikonal method in that relativistic kinematics are easily included. 

The PW solution method is numerically unstable for reactions that have large on-shell momenta, including both high energy reactions and relatively low energy 
AA reactions, due to the highly oscillatory Legendre polynomials needed for convergence of these systems.   
To circumvent this difficulty, the LS3D solution method was implemented. Convergence of the LS3D equation can be achieved quickly after identifying the integration range for momenta and polar angles that give non-zero contributions to the LS equation.
This numerical method is also useful for obtaining convergence for the partial wave analysis; however, numerical instability still exists because of the Legendre polynomial oscillations.   

It was shown that the NR PW and NR LS3D methods agree with the eikonal method, except at very low energies for projectile nucleons, where the eikonal method is not well-suited.  As the lab energy is increased, relativistic effects are seen as a shift in differential cross section resonances toward higher magnitudes and lower angles.  Although some comparisons to experimental data were performed, the aim of this was manuscript was to demonstrate that (1) all three models agree in the appropriate energy regimes (2) there is a noticeable shift in the elastic differential cross section when relativistic kinematics are used (3) the LS3D method can be used for high energy reactions, where PW methods are numerically unstable.  

Based on the results presented herein, it is recommended that the LS3D method be used for high energy NA and AA reactions at all energies because of its rapid numerical convergence and stability. The effect of equal mass kinematics on differential cross sections and extensive comparisons to experimental data will be elucidated in subsequent manuscripts.

\section*{Acknowledgments}
The authors would like to thank Drs. Steve Blattnig, Ryan Norman, Jonathan Ransom, and Francis Badavi for reviewing this manuscript. 
This work was supported by the Human Research Program under the Human Exploration and Operations Mission Directorate of NASA and NASA grant number NNX13AH31A. Khin Maung Maung would like to thank Alexander Maung for his helpful conversations.

%\bibliographystyle{apsrev}
%\bibliography{LS3Dbib}

\begin{thebibliography}{41}
\expandafter\ifx\csname natexlab\endcsname\relax\def\natexlab#1{#1}\fi
\expandafter\ifx\csname bibnamefont\endcsname\relax
  \def\bibnamefont#1{#1}\fi
\expandafter\ifx\csname bibfnamefont\endcsname\relax
  \def\bibfnamefont#1{#1}\fi
\expandafter\ifx\csname citenamefont\endcsname\relax
  \def\citenamefont#1{#1}\fi
\expandafter\ifx\csname url\endcsname\relax
  \def\url#1{\texttt{#1}}\fi
\expandafter\ifx\csname urlprefix\endcsname\relax\def\urlprefix{URL }\fi
\providecommand{\bibinfo}[2]{#2}
\providecommand{\eprint}[2][]{\url{#2}}

\bibitem[{\citenamefont{Ackermann et~al.}(2013)}]{Ackermann}
\bibinfo{author}{\bibfnamefont{M.}~\bibnamefont{Ackermann}}
  \bibnamefont{et~al.}, \bibinfo{journal}{Science}
  \textbf{\bibinfo{volume}{339}}, \bibinfo{pages}{807} (\bibinfo{year}{2013}).

\bibitem[{\citenamefont{Benton and Benton}(2001)}]{Benton}
\bibinfo{author}{\bibfnamefont{E.~R.} \bibnamefont{Benton}} \bibnamefont{and}
  \bibinfo{author}{\bibfnamefont{E.~V.} \bibnamefont{Benton}},
  \bibinfo{journal}{Nucl. Instr. Meth. B} \textbf{\bibinfo{volume}{184}},
  \bibinfo{pages}{255} (\bibinfo{year}{2001}).

\bibitem[{\citenamefont{Slaba et~al.}(2010{\natexlab{a}})\citenamefont{Slaba,
  Blattnig, and Badavi}}]{Slaba1}
\bibinfo{author}{\bibfnamefont{T.~C.} \bibnamefont{Slaba}},
  \bibinfo{author}{\bibfnamefont{S.~R.} \bibnamefont{Blattnig}},
  \bibnamefont{and} \bibinfo{author}{\bibfnamefont{F.~F.}
  \bibnamefont{Badavi}}, \bibinfo{journal}{J. Comput. Phys.}
  \textbf{\bibinfo{volume}{229}}, \bibinfo{pages}{9397}
  (\bibinfo{year}{2010}{\natexlab{a}}).

\bibitem[{\citenamefont{Slaba et~al.}(2010{\natexlab{b}})\citenamefont{Slaba,
  Blattnig, Clowdsley, Walker, and Badavi}}]{Slaba2}
\bibinfo{author}{\bibfnamefont{T.~C.} \bibnamefont{Slaba}},
  \bibinfo{author}{\bibfnamefont{S.~R.} \bibnamefont{Blattnig}},
  \bibinfo{author}{\bibfnamefont{M.~S.} \bibnamefont{Clowdsley}},
  \bibinfo{author}{\bibfnamefont{S.~A.} \bibnamefont{Walker}},
  \bibnamefont{and} \bibinfo{author}{\bibfnamefont{F.~F.}
  \bibnamefont{Badavi}}, \bibinfo{journal}{Adv. Space Res.}
  \textbf{\bibinfo{volume}{46}}, \bibinfo{pages}{800}
  (\bibinfo{year}{2010}{\natexlab{b}}).

\bibitem[{\citenamefont{Wilson et~al.}(1991)\citenamefont{Wilson, Townsend,
  Schimmerling, Khandelwal, Khan, Nealy, Cucinotta, Simonsen, Shinn, and
  Norbury}}]{NASARP1257}
\bibinfo{author}{\bibfnamefont{J.~W.} \bibnamefont{Wilson}},
  \bibinfo{author}{\bibfnamefont{L.~W.} \bibnamefont{Townsend}},
  \bibinfo{author}{\bibfnamefont{W.}~\bibnamefont{Schimmerling}},
  \bibinfo{author}{\bibfnamefont{G.~S.} \bibnamefont{Khandelwal}},
  \bibinfo{author}{\bibfnamefont{F.}~\bibnamefont{Khan}},
  \bibinfo{author}{\bibfnamefont{J.~E.} \bibnamefont{Nealy}},
  \bibinfo{author}{\bibfnamefont{F.~A.} \bibnamefont{Cucinotta}},
  \bibinfo{author}{\bibfnamefont{L.~C.} \bibnamefont{Simonsen}},
  \bibinfo{author}{\bibfnamefont{J.~L.} \bibnamefont{Shinn}}, \bibnamefont{and}
  \bibinfo{author}{\bibfnamefont{J.~W.} \bibnamefont{Norbury}},
  \emph{\bibinfo{title}{Transport methods and interactions for space
  radiations}}, \bibinfo{howpublished}{NASA Reference Publication 1257}
  (\bibinfo{year}{1991}).

\bibitem[{\citenamefont{Simpson}(1983)}]{Simpson}
\bibinfo{author}{\bibfnamefont{J.~A.} \bibnamefont{Simpson}},
  \bibinfo{journal}{Ann. Rev. Nucl. Part. Sci.} \textbf{\bibinfo{volume}{33}},
  \bibinfo{pages}{323} (\bibinfo{year}{1983}).

\bibitem[{\citenamefont{Joachain}(1983)}]{Joachain}
\bibinfo{author}{\bibfnamefont{C.~J.} \bibnamefont{Joachain}},
  \emph{\bibinfo{title}{Quantum Collision Theory}}
  (\bibinfo{publisher}{American Elsevier}, \bibinfo{address}{New York},
  \bibinfo{year}{1983}).

\bibitem[{\citenamefont{Picklesimer
  et~al.}(1984{\natexlab{a}})\citenamefont{Picklesimer, Tandy, Thaler, and
  Wolfe}}]{Pickle}
\bibinfo{author}{\bibfnamefont{A.}~\bibnamefont{Picklesimer}},
  \bibinfo{author}{\bibfnamefont{P.~C.} \bibnamefont{Tandy}},
  \bibinfo{author}{\bibfnamefont{R.}~\bibnamefont{Thaler}}, \bibnamefont{and}
  \bibinfo{author}{\bibfnamefont{D.}~\bibnamefont{Wolfe}},
  \bibinfo{journal}{Phys. Rev. C} \textbf{\bibinfo{volume}{30}},
  \bibinfo{pages}{1861} (\bibinfo{year}{1984}{\natexlab{a}}).

\bibitem[{\citenamefont{Picklesimer
  et~al.}(1984{\natexlab{b}})\citenamefont{Picklesimer, Tandy, Thaler, and
  Wolfe}}]{Pickle2}
\bibinfo{author}{\bibfnamefont{A.}~\bibnamefont{Picklesimer}},
  \bibinfo{author}{\bibfnamefont{P.~C.} \bibnamefont{Tandy}},
  \bibinfo{author}{\bibfnamefont{R.}~\bibnamefont{Thaler}}, \bibnamefont{and}
  \bibinfo{author}{\bibfnamefont{D.}~\bibnamefont{Wolfe}},
  \bibinfo{journal}{Phys. Rev. C} \textbf{\bibinfo{volume}{29}},
  \bibinfo{pages}{1582} (\bibinfo{year}{1984}{\natexlab{b}}).

\bibitem[{\citenamefont{Wolfe}(1983)}]{Wolfe}
\bibinfo{author}{\bibfnamefont{D.~H.} \bibnamefont{Wolfe}}, Ph.D. thesis,
  \bibinfo{school}{Kent State University} (\bibinfo{year}{1983}).

\bibitem[{\citenamefont{{De Jager} et~al.}(1974)\citenamefont{{De Jager}, {De
  Vries}, and {De Vries}}}]{DeVries1}
\bibinfo{author}{\bibfnamefont{C.~W.} \bibnamefont{{De Jager}}},
  \bibinfo{author}{\bibfnamefont{H.}~\bibnamefont{{De Vries}}},
  \bibnamefont{and} \bibinfo{author}{\bibfnamefont{C.}~\bibnamefont{{De
  Vries}}}, \bibinfo{journal}{Atom. Data Nucl. Data}
  \textbf{\bibinfo{volume}{14}}, \bibinfo{pages}{479} (\bibinfo{year}{1974}).

\bibitem[{\citenamefont{{De Vries} et~al.}(1987)\citenamefont{{De Vries}, {De
  Jager}, and {De Vries}}}]{DeVries2}
\bibinfo{author}{\bibfnamefont{H.}~\bibnamefont{{De Vries}}},
  \bibinfo{author}{\bibfnamefont{C.~W.} \bibnamefont{{De Jager}}},
  \bibnamefont{and} \bibinfo{author}{\bibfnamefont{C.}~\bibnamefont{{De
  Vries}}}, \bibinfo{journal}{Atom. Data Nucl. Data}
  \textbf{\bibinfo{volume}{36}}, \bibinfo{pages}{495} (\bibinfo{year}{1987}).

\bibitem[{\citenamefont{Miller et~al.}(2007)\citenamefont{Miller, Reygers,
  Sanders, and Steinberg}}]{Miller}
\bibinfo{author}{\bibfnamefont{M.~L.} \bibnamefont{Miller}},
  \bibinfo{author}{\bibfnamefont{K.}~\bibnamefont{Reygers}},
  \bibinfo{author}{\bibfnamefont{S.~J.} \bibnamefont{Sanders}},
  \bibnamefont{and}
  \bibinfo{author}{\bibfnamefont{P.}~\bibnamefont{Steinberg}},
  \bibinfo{journal}{Annu. Rev. Nucl. Part. Sci.} \textbf{\bibinfo{volume}{57}},
  \bibinfo{pages}{205} (\bibinfo{year}{2007}).

\bibitem[{\citenamefont{Townsend et~al.}(1983)\citenamefont{Townsend,
  Bidasaria, and Wilson}}]{Townsend}
\bibinfo{author}{\bibfnamefont{L.~W.} \bibnamefont{Townsend}},
  \bibinfo{author}{\bibfnamefont{H.~B.} \bibnamefont{Bidasaria}},
  \bibnamefont{and} \bibinfo{author}{\bibfnamefont{J.~W.}
  \bibnamefont{Wilson}}, \bibinfo{journal}{Can. J. Phys.}
  \textbf{\bibinfo{volume}{61}}, \bibinfo{pages}{867} (\bibinfo{year}{1983}).

\bibitem[{\citenamefont{Wilson}(1974)}]{Wilson1974}
\bibinfo{author}{\bibfnamefont{J.~W.} \bibnamefont{Wilson}},
  \bibinfo{journal}{Phys. Lett. B} \textbf{\bibinfo{volume}{52}},
  \bibinfo{pages}{419} (\bibinfo{year}{1974}).

\bibitem[{\citenamefont{Wilson and Townsend}(1981)}]{CJP1981}
\bibinfo{author}{\bibfnamefont{J.~W.} \bibnamefont{Wilson}} \bibnamefont{and}
  \bibinfo{author}{\bibfnamefont{L.~W.} \bibnamefont{Townsend}},
  \bibinfo{journal}{Can. J. Phys.} \textbf{\bibinfo{volume}{59}},
  \bibinfo{pages}{1569} (\bibinfo{year}{1981}).

\bibitem[{\citenamefont{Townsend et~al.}(1982)\citenamefont{Townsend, Wilson,
  and Bidasaria}}]{CJP1982}
\bibinfo{author}{\bibfnamefont{L.~W.} \bibnamefont{Townsend}},
  \bibinfo{author}{\bibfnamefont{J.~W.} \bibnamefont{Wilson}},
  \bibnamefont{and} \bibinfo{author}{\bibfnamefont{H.~B.}
  \bibnamefont{Bidasaria}}, \bibinfo{journal}{Can. J. Phys.}
  \textbf{\bibinfo{volume}{60}}, \bibinfo{pages}{1514} (\bibinfo{year}{1982}).

\bibitem[{\citenamefont{Townsend}(1983)}]{CJP1983}
\bibinfo{author}{\bibfnamefont{L.~W.} \bibnamefont{Townsend}},
  \bibinfo{journal}{Can. J. Phys.} \textbf{\bibinfo{volume}{61}},
  \bibinfo{pages}{93} (\bibinfo{year}{1983}).

\bibitem[{\citenamefont{Werneth et~al.}()\citenamefont{Werneth, Maung, Ford,
  Norbury, and Vera}}]{NASATP2014}
\bibinfo{author}{\bibfnamefont{C.~M.} \bibnamefont{Werneth}},
  \bibinfo{author}{\bibfnamefont{K.~M.} \bibnamefont{Maung}},
  \bibinfo{author}{\bibfnamefont{W.~P.} \bibnamefont{Ford}},
  \bibinfo{author}{\bibfnamefont{J.~W.} \bibnamefont{Norbury}},
  \bibnamefont{and} \bibinfo{author}{\bibfnamefont{M.~D.} \bibnamefont{Vera}},
  \emph{\bibinfo{title}{{Elastic Differential Cross Sections}}},
  \bibinfo{howpublished}{NASA Technical Publication 2014 (submitted)}.

\bibitem[{\citenamefont{Werneth et~al.}(2013)\citenamefont{Werneth, Maung,
  Mead, and Blattnig}}]{Werneth_Finite}
\bibinfo{author}{\bibfnamefont{C.~M.} \bibnamefont{Werneth}},
  \bibinfo{author}{\bibfnamefont{K.~M.} \bibnamefont{Maung}},
  \bibinfo{author}{\bibfnamefont{L.~R.} \bibnamefont{Mead}}, \bibnamefont{and}
  \bibinfo{author}{\bibfnamefont{S.~R.} \bibnamefont{Blattnig}},
  \bibinfo{journal}{Nucl. Instr. Meth. B} \textbf{\bibinfo{volume}{308}},
  \bibinfo{pages}{40} (\bibinfo{year}{2013}).

\bibitem[{\citenamefont{Elster et~al.}(1997{\natexlab{a}})\citenamefont{Elster,
  Thomas, and Glockle}}]{LS3D_1}
\bibinfo{author}{\bibfnamefont{C.}~\bibnamefont{Elster}},
  \bibinfo{author}{\bibfnamefont{J.~H.} \bibnamefont{Thomas}},
  \bibnamefont{and} \bibinfo{author}{\bibfnamefont{W.}~\bibnamefont{Glockle}},
  \bibinfo{howpublished}{arXiv:9708017v1 [nucl-th]}
  (\bibinfo{year}{1997}{\natexlab{a}}).

\bibitem[{\citenamefont{Fachruddin et~al.}(2000)\citenamefont{Fachruddin,
  Elster, and Glockle}}]{LS3D_2}
\bibinfo{author}{\bibfnamefont{I.}~\bibnamefont{Fachruddin}},
  \bibinfo{author}{\bibfnamefont{C.}~\bibnamefont{Elster}}, \bibnamefont{and}
  \bibinfo{author}{\bibfnamefont{W.}~\bibnamefont{Glockle}},
  \bibinfo{journal}{Phys. Rev. C} \textbf{\bibinfo{volume}{62}},
  \bibinfo{pages}{044002} (\bibinfo{year}{2000}).

\bibitem[{\citenamefont{Rodriguez-Gallardo
  et~al.}(2008)\citenamefont{Rodriguez-Gallardo, Deltuva, Cravo, Crespo, and
  Fonseca}}]{LS3D_3}
\bibinfo{author}{\bibfnamefont{M.}~\bibnamefont{Rodriguez-Gallardo}},
  \bibinfo{author}{\bibfnamefont{A.}~\bibnamefont{Deltuva}},
  \bibinfo{author}{\bibfnamefont{E.}~\bibnamefont{Cravo}},
  \bibinfo{author}{\bibfnamefont{R.}~\bibnamefont{Crespo}}, \bibnamefont{and}
  \bibinfo{author}{\bibfnamefont{A.~C.} \bibnamefont{Fonseca}},
  \bibinfo{journal}{Phys. Rev. C} \textbf{\bibinfo{volume}{78}},
  \bibinfo{pages}{034602} (\bibinfo{year}{2008}).

\bibitem[{\citenamefont{Veerasamy et~al.}(2013)\citenamefont{Veerasamy, Elster,
  and Polyzou}}]{LS3D_4}
\bibinfo{author}{\bibfnamefont{S.}~\bibnamefont{Veerasamy}},
  \bibinfo{author}{\bibfnamefont{C.}~\bibnamefont{Elster}}, \bibnamefont{and}
  \bibinfo{author}{\bibfnamefont{W.~N.} \bibnamefont{Polyzou}},
  \bibinfo{journal}{Few Body Systems} \textbf{\bibinfo{volume}{54}},
  \bibinfo{pages}{2207} (\bibinfo{year}{2013}).

\bibitem[{\citenamefont{Liu et~al.}(2005)\citenamefont{Liu, Elster, and
  Glockle}}]{LS3D_3body}
\bibinfo{author}{\bibfnamefont{H.}~\bibnamefont{Liu}},
  \bibinfo{author}{\bibfnamefont{C.}~\bibnamefont{Elster}}, \bibnamefont{and}
  \bibinfo{author}{\bibfnamefont{W.}~\bibnamefont{Glockle}},
  \bibinfo{journal}{Phys. Rev. C} \textbf{\bibinfo{volume}{72}},
  \bibinfo{pages}{054003} (\bibinfo{year}{2005}).

\bibitem[{\citenamefont{Maung et~al.}(2007)\citenamefont{Maung, Norbury, and
  Coleman}}]{Maung_Norbury_RMST}
\bibinfo{author}{\bibfnamefont{K.~M.} \bibnamefont{Maung}},
  \bibinfo{author}{\bibfnamefont{J.~W.} \bibnamefont{Norbury}},
  \bibnamefont{and} \bibinfo{author}{\bibfnamefont{T.}~\bibnamefont{Coleman}},
  \bibinfo{journal}{J. Phys. G: Nucl. Part. Phys.}
  \textbf{\bibinfo{volume}{34}}, \bibinfo{pages}{1861} (\bibinfo{year}{2007}).

\bibitem[{\citenamefont{Feshbach}(1958)}]{Feshbach1}
\bibinfo{author}{\bibfnamefont{H.}~\bibnamefont{Feshbach}},
  \bibinfo{journal}{Ann. Phys.} \textbf{\bibinfo{volume}{5}},
  \bibinfo{pages}{357} (\bibinfo{year}{1958}).

\bibitem[{\citenamefont{Feshbach}(1962)}]{Feshbach2}
\bibinfo{author}{\bibfnamefont{H.}~\bibnamefont{Feshbach}},
  \bibinfo{journal}{Ann. Phys.} \textbf{\bibinfo{volume}{19}},
  \bibinfo{pages}{287} (\bibinfo{year}{1962}).

\bibitem[{\citenamefont{Watson}(1953)}]{Watson}
\bibinfo{author}{\bibfnamefont{K.~M.} \bibnamefont{Watson}},
  \bibinfo{journal}{Phys. Rev.} \textbf{\bibinfo{volume}{89}},
  \bibinfo{pages}{575} (\bibinfo{year}{1953}).

\bibitem[{\citenamefont{Werneth and Maung}(2013)}]{Werneth_RMST}
\bibinfo{author}{\bibfnamefont{C.~M.} \bibnamefont{Werneth}} \bibnamefont{and}
  \bibinfo{author}{\bibfnamefont{K.~M.} \bibnamefont{Maung}},
  \bibinfo{journal}{Can. J. Phys.} \textbf{\bibinfo{volume}{91}},
  \bibinfo{pages}{424} (\bibinfo{year}{2013}).

\bibitem[{\citenamefont{Elster et~al.}(1997{\natexlab{b}})\citenamefont{Elster,
  Weppner, and Chinn}}]{Elster}
\bibinfo{author}{\bibfnamefont{C.}~\bibnamefont{Elster}},
  \bibinfo{author}{\bibfnamefont{S.~P.} \bibnamefont{Weppner}},
  \bibnamefont{and} \bibinfo{author}{\bibfnamefont{C.~R.} \bibnamefont{Chinn}},
  \bibinfo{journal}{Phys. Rev. C} \textbf{\bibinfo{volume}{56}},
  \bibinfo{pages}{2080} (\bibinfo{year}{1997}{\natexlab{b}}).

\bibitem[{\citenamefont{M\"{o}ller}(1945)}]{Moller}
\bibinfo{author}{\bibfnamefont{C.}~\bibnamefont{M\"{o}ller}},
  \bibinfo{journal}{K. Dan. Vidensk. Selsk. Mat.-Fys. Medd.}
  \textbf{\bibinfo{volume}{23}}, \bibinfo{pages}{1} (\bibinfo{year}{1945}).

\bibitem[{\citenamefont{Landau}(1996)}]{QMII}
\bibinfo{author}{\bibfnamefont{R.~H.} \bibnamefont{Landau}},
  \emph{\bibinfo{title}{{Quantum Mechanics II: A Second Course in Quantum
  Theory}}} (\bibinfo{publisher}{{John Wiley \& Sons Inc.}},
  \bibinfo{address}{New York}, \bibinfo{year}{1996}).

\bibitem[{\citenamefont{Sloan}(1968)}]{Sloan}
\bibinfo{author}{\bibfnamefont{I.~H.} \bibnamefont{Sloan}},
  \bibinfo{journal}{J. Comput. Phys.} \textbf{\bibinfo{volume}{3}},
  \bibinfo{pages}{332} (\bibinfo{year}{1968}).

\bibitem[{\citenamefont{Glauber}(1959)}]{Glauber}
\bibinfo{author}{\bibfnamefont{R.~J.} \bibnamefont{Glauber}},
  \emph{\bibinfo{title}{Lectures in Theoretical Physics}}
  (\bibinfo{publisher}{Interscience Publishers Inc.}, \bibinfo{address}{New
  York}, \bibinfo{year}{1959}).

\bibitem[{\citenamefont{Meyers and Swiatecki}(1974)}]{Myers}
\bibinfo{author}{\bibfnamefont{W.~D.} \bibnamefont{Meyers}} \bibnamefont{and}
  \bibinfo{author}{\bibfnamefont{W.~J.} \bibnamefont{Swiatecki}},
  \bibinfo{journal}{Ann. Phys.} \textbf{\bibinfo{volume}{84}},
  \bibinfo{pages}{186} (\bibinfo{year}{1974}).

\bibitem[{\citenamefont{Norbury}()}]{NASATP}
\bibinfo{author}{\bibfnamefont{J.~W.} \bibnamefont{Norbury}},
  \emph{\bibinfo{title}{{Total Nucleon-Nucleon Cross Section}}},
  \bibinfo{howpublished}{NASA Technical Publication 2008-215116}.

\bibitem[{\citenamefont{Alkhazov et~al.}(1972)\citenamefont{Alkhazov, Amalsky,
  Belostotsky, Vorobyov, Domchenkov, and Dotsenko}}]{pS32}
\bibinfo{author}{\bibfnamefont{G.~D.} \bibnamefont{Alkhazov}},
  \bibinfo{author}{\bibfnamefont{G.~M.} \bibnamefont{Amalsky}},
  \bibinfo{author}{\bibfnamefont{S.~L.} \bibnamefont{Belostotsky}},
  \bibinfo{author}{\bibfnamefont{A.~A.} \bibnamefont{Vorobyov}},
  \bibinfo{author}{\bibfnamefont{O.~A.} \bibnamefont{Domchenkov}},
  \bibnamefont{and} \bibinfo{author}{\bibfnamefont{Y.~V.}
  \bibnamefont{Dotsenko}}, \bibinfo{journal}{Phys. Lett. B}
  \textbf{\bibinfo{volume}{42}}, \bibinfo{pages}{121} (\bibinfo{year}{1972}).

\bibitem[{\citenamefont{Hoffmann et~al.}(1981)}]{pCa40}
\bibinfo{author}{\bibfnamefont{G.~W.} \bibnamefont{Hoffmann}}
  \bibnamefont{et~al.}, \bibinfo{journal}{Phys. Rev. Lett.}
  \textbf{\bibinfo{volume}{47}}, \bibinfo{pages}{1436} (\bibinfo{year}{1981}).

\bibitem[{\citenamefont{Lombard et~al.}(1981)\citenamefont{Lombard, Alkhazov,
  and Domchenkov}}]{pNi58}
\bibinfo{author}{\bibfnamefont{R.~M.} \bibnamefont{Lombard}},
  \bibinfo{author}{\bibfnamefont{G.~D.} \bibnamefont{Alkhazov}},
  \bibnamefont{and} \bibinfo{author}{\bibfnamefont{O.~A.}
  \bibnamefont{Domchenkov}}, \bibinfo{journal}{Nucl. Phys. A}
  \textbf{\bibinfo{volume}{360}}, \bibinfo{pages}{233} (\bibinfo{year}{1981}).

\bibitem[{\citenamefont{Alkhazov et~al.}(1977)}]{HeCa40}
\bibinfo{author}{\bibfnamefont{G.~D.} \bibnamefont{Alkhazov}}
  \bibnamefont{et~al.}, \bibinfo{journal}{Nucl. Phys. A}
  \textbf{\bibinfo{volume}{280}}, \bibinfo{pages}{365} (\bibinfo{year}{1977}).

\end{thebibliography}

\newpage

\clearpage

\clearpage

\end{document}